\documentclass{ws-rv9x6}       
\usepackage{epic} 

\begin{document}

\setcounter{chapter}{0}

\chapter{One-Dimensional Quantum Spin Liquids}

\author{P. Lecheminant}

\address{Laboratoire de Physique Th\'eorique et Mod\'elisation, 
CNRS UMR 8089, \\
Universit{\'e} de Cergy-Pontoise,
5 mail Gay-Lussac, Neuville sur Oise, \\
95301 Cergy-Pontoise Cedex, France}

\begin{abstract}
This chapter is intended as a brief overview of some of the
quantum spin liquid phases with unbroken SU(2) spin symmetry 
available in one dimension.
The main characteristics of these phases are discussed
by means of the bosonization approach.  
A special emphasis is laid on 
the interplay between frustration and quantum fluctuations
in one dimension.
\end{abstract}

\section{Introduction} 

A central issue in the study of strongly correlated
systems is the classification of all possible 
Mott insulating phases at zero temperature.
The general strategy for describing the possible phases associated with the 
spin degrees of freedom (the so-called ``quantum magnetism'') of
Mott insulators consists of analysing localized spin models
such as the antiferromagnetic (AF) Heisenberg model:
\begin{equation}
{\cal H} = J\sum_{<i,j>} {\vec S}_i \cdot {\vec S}_j +..,
\label{genheis}
\end{equation}
where ${\vec S}_i$ are quantum spin-$S$ operators on sites
$i$ of a lattice, and the sum $<i,j>$ is over
the nearest neighbor sites with an
antiferromagnetic exchange ($J > 0$);
the ellipsis represents additional terms like second-neighbor
competing exchange interaction or ring exchange
that can be eventually added to the Heisenberg Hamiltonian (\ref{genheis}).
Generically, the ground state of the model (\ref{genheis})
without these additional terms displays long-range N{\'e}el ordering
($<{\vec S}_i> \ne \vec 0$) which breaks spontaneously the SU(2)
symmetry of the lattice Hamiltonian (\ref{genheis}).
The low-energy excitations are gapless spin-waves (i.e. magnons)
as expected when a continuous symmetry is spontaneously broken.

A central focus of quantum magnetism over the years, starting from 
the proposal made by Anderson \cite{anderson73},
has been the search for a spin liquid behavior i.e. 
a phase with no magnetic long-range N{\'e}el order.
Spin liquids phases are expected to be stabilized in low
dimensions or in presence of frustration, 
i.e. in situations where quantum fluctuations can strongly suppress 
magnetism. It gives rise to rich physics 
with exotic low-energy excitations
which require in most cases the use of non-perturbative 
techniques to fully determine its properties. 
In the two-dimensional case, different spin liquid phases 
have been found and display bond ordering 
or topological ordering
(for a review see for instance 
Refs. \cite{sachdev02,lhuillier01} and references therein).
An interesting attemp to develop a systematic classification of two-dimensional
spin liquids has also recently been explored by Wen \cite{wen01}.

In one dimension, the quantum analog of the Mermin-Wagner \cite{mermin68}
theorem shows that quantum fluctuations always 
disorder the ground state in systems with a continuous
symmetry. The generic situation in one dimension is thus 
to have a spin liquid phase.
The one dimensional case is also extremely  favorable 
since several non-perturbative techniques are available
to fully characterize the physical properties of different
spin liquids.
These powerful techniques include
integrability, conformal field theory (CFT) \cite{bpz,bookcft}, 
the bosonization approach\cite{luther74,coleman75,bookboso}, 
and numerical calculations such as the density matrix 
renormalization group (DMRG) approach \cite{white92}.
The determination of the 
properties of spin liquids is not a purely academic problem since
many physical realizations of spin chains have been
synthesized over the years.
The reader may consult for instance the recent reviews
\cite{yamashita00,lemmens03}
on experiments in quasi-one dimensional spin systems.

The main questions in the study of one-dimensional
spin liquids are in order:
What are the different spin liquid phases available 
at zero-temperature? 
What are the different physical properties of these spin liquids?
What is the nature of the quantum phase transition \cite{sachdevbook}
between two different spin liquids?
Last but not least, the problem of dimensional cross-over,
i.e. the fate of the spin liquid as the two dimensional case is
reached is also a central topic. In particular, is it
possible to stabilize a spin liquid phase in $d=2$ starting from 
the $d=1$ case? 

In this work, we shall mainly be concerned with the two first questions
for spin liquids with unbroken SU(2) spin symmetry.
Using the bosonization approach, we shall discuss
some of the physical properties of
spin liquids that have been found in spin chains, 
spin ladders with or without frustration.
An important question related to this problem is 
how can one classify all these 1D spin liquid phases at $T=0$?
A first distinction concerns the existence or absence of a spectral
gap in the model. In a gapless system (critical spin liquid), one 
has quasi-long range N{\'e}el order.
The leading asymptotics of spin-spin correlation functions 
display a power law behavior with an exponent characterized
by a well defined underlying CFT \cite{affleck85}.
In contrast, in a spin liquid phase with a spectral gap,
spin-spin correlation functions have an exponential decay
with the distance due to the existence of a finite correlation
length $\xi$: 
$\langle {\vec S}_i \cdot {\vec S}_j \rangle \sim \exp(- |i-j|/\xi)$.
At this point, it is worth stressing that two gapful spin liquid 
phases sharing the same thermodynamic properties (typically a thermal 
activation law at low T) may display very different behaviors in other physical 
quantities of interest such as, for instance, dynamical or optical properties.
It is thus necessary to make a scrutiny analysis 
of the properties of the phase before elaborating a complete classification. 
Let us first consider the
ground state of a gapful spin liquid phase.
A (discrete) symmetry might be spontaneously broken in the 
ground state resulting on a ground-state degeneracy.
The spin liquid phase may display also an hidden topological order \cite{wen95} 
which manifests itself in a ground-state degeneracy which depends 
on the nature of boundary conditions (BC) used i.e. periodic or open BC.
The resulting chain-end (edge) excitations in the open BC case can 
be observed in the NMR profile of a spin chain compound doped with
non-magnetic impurities like Zn or Mg.
Another important distinction between two spin liquids relies
on the quantum number carried by elementary excitations.
The fractionalized or integer nature of this quantum number
has crucial consequences in the dynamical structure factor
of the system which directly probes the elementary nature
of the spin-flip.
A sharp spectral peak, characteristic of 
a well defined $S=1$ mode, or a broad (incoherent background) feature 
in the dynamical structure factor can be seen in inelastic 
neutron scattering experiments.
Finally, an interesting question is the existence or not of
bound states in the energy spectrum below the two-particle continuum.
Spin liquid phases may then be distinguished at this level
which is also an important fact from the experimental point of view
since observation of magnetic singlet bound states can 
be realized through light scattering experiments \cite{lemmens03}.
This classification is certainly far from being complete
but it enables us to distinguish between several 
spin liquid phases and investigate the nature of the 
quantum phase transition between them.

This chapter is organized as follows:
We present, in Section 2, the different spin liquid phases
that occur in unfrustrated spin chains and spin ladders.
The main effects of frustration in one-dimensional
spin liquids are described in Section 3.
In particular,
we shall observe that frustration 
plays its trick 
by allowing deconfined spinons (carrying fractional $S=1/2$ quantum number) 
as elementary excitations
and it provides a non-trivial source of incommensurability.
Finally, our concluding remarks are given in Section 4.

\section{Unfrustrated spin chains}     
The paradigmatic model to investigate the properties
of 1D quantum magnets in absence of frustration 
is the AF Heisenberg spin chain given by the Hamiltonian:
\begin{equation}
{\cal H} = J\sum_{i} {\vec S}_i \cdot {\vec S}_{i+1},
\label{heischain}
\end{equation}
where ${\vec S}_i$ is a spin-$S$ operator at the ith site of the chain
and the exchange interaction is antiferromagnetic: $J>0$.
In the classical limit, the model displays N{\'e}el long-range 
order: ${\vec S}_i \sim S (-1)^i {\vec n}$, ${\vec n}$
being an unit vector with an arbitrary fixed orientation in spin space.
This solution breaks the SU(2) symmetry of the model (\ref{heischain})
down to U(1). Corresponding to this symmetry breaking scheme, 
there are two Goldstone modes which propagate
with the same velocity (Lorentz invariance in the low-energy limit).
In the language of CFT \cite{bookcft}, it corresponds to a 
field theory with central charge $c=2$ ($c=1$
being the central charge of a free massless boson field).
Physically, these Goldstone modes are nothing but a doublet of gapless
spin waves modes associated with slow modulations in the
orientation of the vector ${\vec n}$ which represents
the order parameter of the N{\'e}el magnetic structure.
In the quantum case, this N{\'e}el solution in one dimension 
is destabilized by strong quantum fluctuations no matter
how large $S$ is: a spin liquid phase is formed.

\subsection{Spin-1/2 Heisenberg chain}
In the ultra quantum limit i.e. $S=1/2$, the nature of the
spin liquid phase can be fully determined since the 
model (\ref{heischain}) is exactly solvable for $S=1/2$
by means of the Bethe ansatz approach.
In particular, starting with Bethe's seminal work \cite{bethe31},
a host of exact results have been obtained 
over the years for ground state
properties \cite{hulthen38}, magnetic susceptibility \cite{suscept},
thermodynamics \cite{thermo}, excitation spectrum \cite{dcp62,faddeev81}, 
and correlation functions \cite{correl}. 

The model displays quantum criticality properties which 
belong to the Wess-Zumino-Novikov-Witten (WZNW) su(2)$_1$ 
universality class \cite{affleck85}. The low-energy limit
of the spin-1/2 AF Heisenberg chain (\ref{heischain})
is described by the su(2)$_1$ WZNW CFT with central charge
$c=1$ perturbed by marginally irrelevant
current-current interaction \cite{affleck85,affleckhouches}.
The resulting Hamiltonian density reads as follows:
\begin{equation}
{\cal H}_{\rm eff} = \frac{2\pi v}{3} \left({\vec J}_L^2 + 
{\vec J}_R^2 \right) +\lambda {\vec J}_R \cdot {\vec J}_L,
\label{heiscont}
\end{equation}
where ${\vec J}_R$ and ${\vec J}_L$ are respectively the 
right and left su(2)$_1$ currents which generate the 
su(2)$_1$ WZNW CFT \cite{kz84}; these currents satisfy the 
su(2)$_1$ Kac-Moody commutation relations:
\begin{equation}
\left[J^a_{R,L}\left(x\right), J^b_{R,L}\left(y\right) \right] = 
\mp \frac{i \delta^{ab}}{4\pi} \delta^{'}\left(x-y\right) 
+ i \epsilon^{abc} J^c_{R,L}\left(x\right) \delta\left(x-y\right).
\label{kacmoodycomut}
\end{equation}
In Eq. (\ref{heiscont}), $v$ is the spin velocity and
$\lambda <0$ so that the last contribution is a 
marginally irrelevant term that renormalizes to zero in 
the far infrared (IR) limit.
This perturbation accounts for logarithmic 
corrections in the spin-spin correlation \cite{log} which 
is exactly known \cite{affleck98}
in the long-distance limit: 
\begin{equation}
\langle {\vec S}_0 \cdot {\vec S}_r \rangle {\simeq}_{r \gg 1} 
\frac{\left(-1\right)^r}{\left(2\pi\right)^{3/2}} 
\frac{\left(\ln r\right)^{1/2}}{r} .
\label{correlhei}
\end{equation}
 
\begin{figure}[ht]
\begin{center}
\noindent
\epsfxsize=0.5\textwidth
\epsfbox{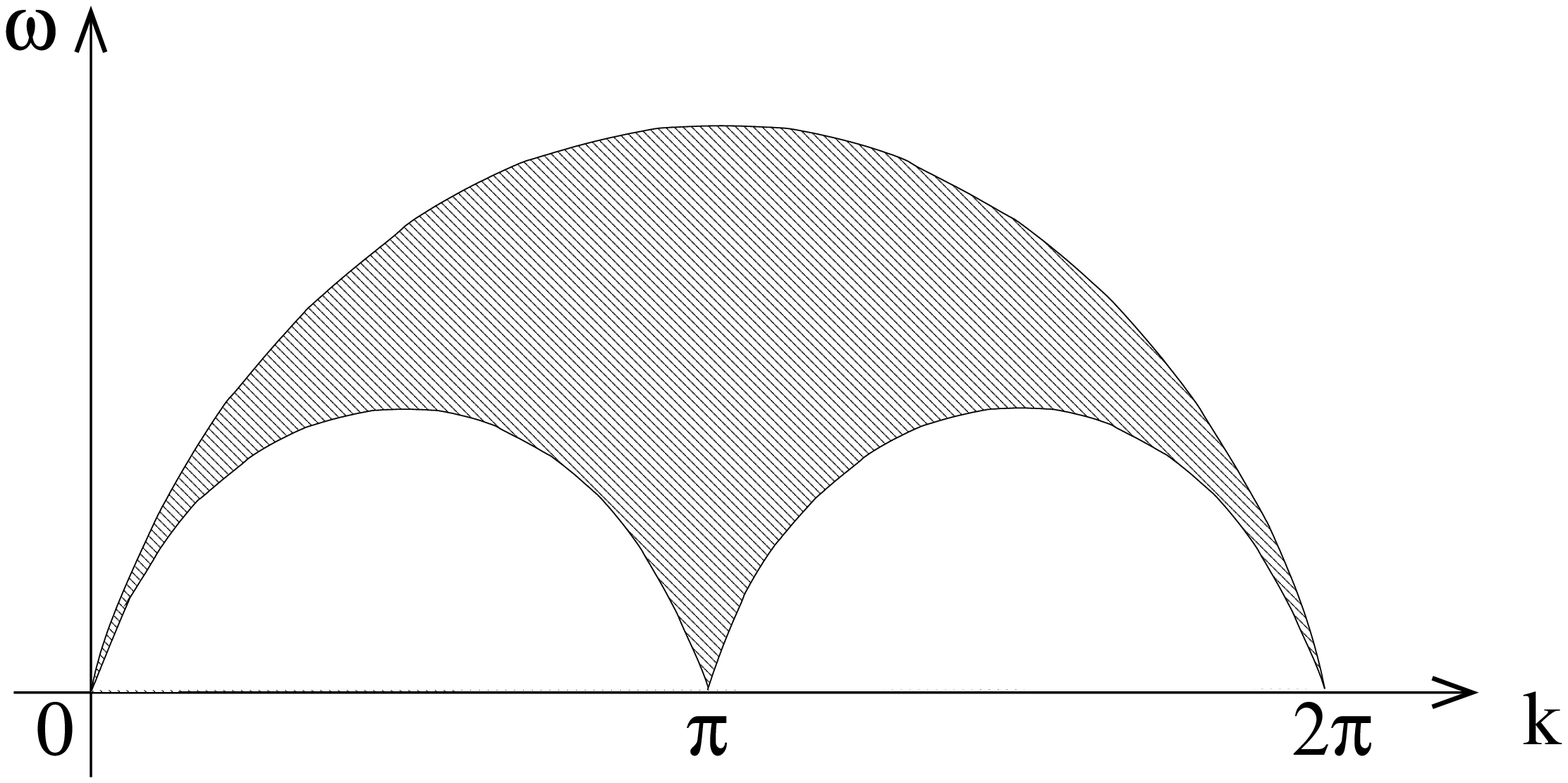}
\end{center}
\caption{\label{figspinon}%
Two-spinon continuum of the spin-1/2 AF Heisenberg chain.
}
\end{figure}

The most striking feature of this spin liquid phase 
stems from the nature of its elementary excitations.
They have been elucidated by Faddeev
and Takhtajan \cite{faddeev81} within the Bethe
ansatz approach and consist of fractional $S=1/2$
massless excitations called spinons. 
The lowest excitations are fourthfold degenerate 
and correspond to a triplet ($S=1$) and a singlet
($S=0$). The resulting energy spectrum is a continuum
in (k,$\omega$) space between a lower boundary (the 
des Cloizeaux-Pearson dispersion relation \cite{dcp62})
$\omega_{\rm dcp} = \pi J |\sin k|/2$ ($-\pi < k \le \pi$)
and an upper boundary $\omega_{u} = \pi J |\sin (k/2)|$
(see Fig. \ref{figspinon}).
The central point of the analysis is that this continuum
can be interpreted as being made up of two spin-1/2 
excitations (spinons) with the dispersion:
\begin{equation}
\omega_{\rm spinon} = \frac{\pi J}{2} \sin k ,
\label{spinondisp}
\end{equation}
with $0 < k < \pi$. 
A spinon has thus a wave-vector restricted to
only half of the Brillouin zone.
A triplet (or singlet) excitation with momentum k 
is then described by two spinons with momenta $k_1$ 
and $k_2$ ($0 < k_1 \le k_2 < \pi$) such that:
$\omega(k) = \omega_{\rm spinon}(k_1) + 
\omega_{\rm spinon}(k_2)$ with $k=k_1 + k_2$ 
(respectively $k=k_1 + k_2 - 2 \pi$) if $0 < k \le \pi$
(respectively $-\pi < k < 0$).
It corresponds to a two-parameter continuum
$\omega(k) = \pi J \sin(k/2) \cos(k/2 - q_1)$
with $0 < q_1 < k/2$ for $0 < k < \pi$ 
and $\pi + k  < q_1 < \pi + k/2$ for $-\pi < k < 0$ 
which identifies with the continuum of the spin-1/2 AF Heisenberg
chain.

A magnon excitation carrying $S=1$ quantum number 
is thus no longer an elementary excitation in this model
but is fractionalized into two spinons. 
One important experimental signature of these spinons
is the absence of any sharp peak in the dynamical susceptibility
structure factor since a single spin-flip generates 
a triplet excitation which is not elementary here but made up
of two spinons. The spectral density 
is then a convolution of the spectral densities of the individual
spinons. These spinons have fascinating properties
in particular they obey semion statistics
intermediate between bosons and fermions \cite{haldane91}
as it can be seen by the study of the elementary excitations
of the Haldane-Shastry model \cite{hsmodel} which belongs
to the same universality class as the spin-1/2 AF Heisenberg chain.
An heuristic way to describe the spinons is to consider 
an anisotropic XXZ version of the Heisenberg model (\ref{heischain}):
\begin{equation}
{\cal H}_{\rm XXZ} = J_z \sum_i S^z_i S^z_{i+1}
+ \frac{J_{\perp}}{2} \sum_i \left( S^+_i S^-_{i+1}
+ S^+_{i+1} S^-_{i}\right).
\label{xxzham}
\end{equation}       
This model is still integrable and the ground state is two-fold
degenerate for  $J_z > J_{\perp}$ as the result of the spontaneous
breaking of a discrete Z$_2$ symmetry. 
In this regime, a spectral gap is opened and the elementary excitations
are massive spinons.
A simple way to understand these properties is to consider
the Ising limit when $J_z \gg J_{\perp}$ where the ground state
reduces to the two N{\'e}el ordered states:
$\big \uparrow \big \downarrow \big \uparrow 
\big  \downarrow
\big  \uparrow \big  \downarrow .. $
and $\big \downarrow \big \uparrow \big \downarrow \big \uparrow
\big \downarrow \big \uparrow ..$.
The elementary excitations are massive kinks which carry 
$\Delta S^z= \pm 1/2$ quantum number and interpolate 
between these two ground states:
$\big \uparrow  \big \downarrow  \big \uparrow 
\big \downarrow 
\big \uparrow  \times   \big \uparrow 
\big \downarrow  \big \uparrow..$
The composite nature of a magnon excitation can then be readily 
seen by starting from a ground state and apply a single
spin-flip on it to obtain the configuration:
$\big \uparrow \big \downarrow \big \uparrow 
\big \downarrow \big \downarrow \big \downarrow \big \uparrow 
\big \downarrow 
\big \uparrow \big \downarrow$.
Applying the exchange $J_{\perp}$ interaction 
of the model (\ref{xxzham}) on this state, this magnon excitation 
with $\Delta S^z = -1$ is in fact made up 
of two kinks of the N{\'e}el order i.e. two spinons:
$\big \uparrow \big \downarrow \times \big \downarrow 
\big \uparrow \big \downarrow \big \uparrow \big \downarrow \times
\big \downarrow \big \uparrow \big \downarrow$.
At the transition $J_z = J_{\perp}$ (isotropic limit), the ground state is 
disordered by ``condensation'' of these kinks and 
the low-lying excitations carrying $\Delta S^z = \pm 1/2$ 
are transmuted from gapful kinks to gapless spinons.
Magnons are still built as pairs of these spinons
and have a gapless energy spectrum.

This remarkable production of two particles from 
a single spin-flip has been explored in inelastic
neutron scattering measurements on quasi-1D compounds
that are good experimental realizations of the model (\ref{heischain}).
The first clear evidence of spinons comes from neutron 
experiments \cite{nagler91} on KCuF$_3$ that found a continuum of magnetic
excitations consistent with that expected from 
unbound spinons pair.
The spinon continuum has also been observed in several other 
realizations of the spin-1/2 AF Heisenberg chain 
as the copper benzoate Cu(C$_6$D$_5$COO)$_2$.3D$_2$O \cite{dender96},
BaCu$_2$Si$_2$O$_7$ \cite{tsukada99}, and 
Cu(C$_4$H$_4$N$_2$)(NO$_3$)$_2$ \cite{hammar99}.

\subsection{Haldane's conjecture}

After the elucidation of the elementary 
excitations of the spin-1/2 AF Heisenberg chain
by Faddeev and Takhtajan \cite{faddeev81} in the early eighties,
a central question of 1D quantum magnetism concerns 
the stability of the spinons.
What is the fate of these exotic excitations upon 
switching on small deviations from the model (\ref{heischain})?
Do these additional terms lead to a confinement of the spinons or 
do they still remain deconfined?

A simple modification of the spin-1/2 AF Heisenberg chain (\ref{heischain})
is not to add perturbations but to change the Hilbert 
space namely to consider the general spin-$S$ case.
The resulting model is no longer integrable and one has to resort to
approximate methods to describe its physical properties. 
This question has lead to one of the most profound result
in the field of 1D quantum magnetism: the difference
between integer and half-integer spins AF Heisenberg chains 
predicted by Haldane \cite{haldane83}.
Integer spin chains are incompressible spin liquids with 
a finite gap (the so-called Haldane gap) in the energy spectrum.
In that case, the spinons are confined and 
the elementary excitations of the model are massive (optical) $S=1$ magnons.
In contrast, for half-integer spins, 
the model displays quantum criticality 
with similar universal properties as the $S=1/2$ case
that are described by the su(2)$_1$ WZNW universality class.
The elementary excitations are still, in this case, 
gapless spinons.

This remarkable distinction between integer and half-integer spins,
historically called Haldane's conjecture, has been obtained
by means of a semiclassical analysis of the model with a special 
emphasis on the topological nature of the order parameter
fluctuations. For a review of this approach,
the reader may consult the reviews \cite{affleckrevue,affleckhouches}
and the books \cite{fradkinbook,auerbachbook,sachdevbook}.
In the large $S$ limit, the Euclidean action
that describes the low-energy properties of the spin-$S$ AF Heisenberg
chain reads as follows
\begin{equation}
{\cal S}_{\rm eff} = \frac{v}{2g} \int d\tau dx \left[
\frac{1}{v^2} \left(\partial_{\tau} {\vec n}\right)^2
+ \left(\partial_{x} {\vec n}\right)^2 \right]
+ i \; 2 \pi S {\cal Q}\left({\vec n}\right),
\label{actionsigma}
\end{equation}
where $v = 2 J S$, $g = 2/S$, and ${\vec n}$ is
the order parameter of the N{\'e}el collinear state.
The second contribution in Eq. (\ref{actionsigma}) 
is a topological term since
${\cal Q}\left({\vec n}\right) = \int d\tau dx
\; {\vec n} \cdot (\partial_{\tau} {\vec n} \wedge
\partial_{x} {\vec n})/4\pi$
is an integer, called the Pontryagin index, 
which measures the number of times
the spin configuration ${\vec n}(x,\tau)$ covers the 
surface of the unit sphere S$^2$.
In more mathematical terms, the configurations ${\vec n}(x,\tau)$,
with fixed boundary conditions at infinity,
are mappings of the sphere S$^2$ onto S$^2$ with homotopy classes
classified by an integer $\Pi_2(S^2) = Z$, which is nothing 
but the Pontryagin index $Q$.
For integer spin chains, the term $2 \pi S {\cal Q}\left({\vec n}\right)$
in Eq. (\ref{actionsigma}) has no effect on the path integral of the model 
and can be discarded.
The effective action (\ref{actionsigma}) reduces then to
the one of the two-dimensional O(3) non-linear sigma model
which is a massive integrable field theory \cite{zamolo79,wiegmann85}.
The exact low-energy spectrum of this field theory consists of 
a massive bosonic triplet with mass $m \sim e^{- 2\pi/g}$
and it exhibits no bound states.
The AF Heisenberg chain with integer spin is thus expected
to have low-lying triplet excitations
with a gap that scales as $\Delta \sim v e^{-\pi S}$
when $S$ is large. 

In the case of half-integer spins, the topological term in 
Eq. (\ref{actionsigma}) manifests itself in the path integral 
through a phase factor $(-1)^{\cal Q}$ which gives 
rise to quantum interference between topologically 
distinct paths in space-time of the order parameter 
field ${\vec n}(x,\tau)$.
It turns out that this process protects the model from
a dynamically generated mass gap by quantum fluctuations
and, in contrast, it leads to a non-perturbative massless flow towards
an IR conformally invariant 
fixed point \cite{affleck87,shankar90,zamolo92}
which belongs to the su(2)$_1$ WZNW universality class.
Since this fixed point describes also the universal properties 
of the spin-1/2 AF Heisenberg chain, all half-integer spins 
chains should exhibit the same type of emerging quantum criticality
with for instance spin-spin correlations that decay according
to the power law behavior (\ref{correlhei}).  

\subsection{Haldane spin liquid: spin-1 Heisenberg chain}

The simplest incompressible one-dimensional spin liquid phase
corresponds to the spin-1 AF Heisenberg chain.
The existence of a Haldane gap in this model has generated
an intense activity over the years after its prediction.
It has been confirmed numerically from 
exact diagonalizations on finite samples \cite{exactdiag}, 
quantum Monte-Carlo methods \cite{qmc},  
transfert matrix computations \cite{transfert},
and finally from DMRG calculations \cite{white92,dmrg}.
In particular, this latter technique predicts 
a gap $\Delta = 0.41050(2) J$
and a correlation length $\xi \simeq 6.03 (1)$ 
lattice spacings.                    
From the experimental point of view, several Haldane compounds 
have been synthesized over the years (see for instance
Ref. \cite{yamashita00} for a recent review).
The two most studied compounds are CsNiCl$_3$ \cite{csnicl3} and
Ni(C$_2$H$_8$N$_2$)$_2$NO$_2$(ClO$_4$) (NENP) \cite{renard87}
where spin-1 local moments are provided by Ni$^{2+}$ ions.  
In particular, inelastic neutron scattering experiments 
on NENP \cite{renard87,neunenp} confirm the existence 
of the Haldane gap.
This energy gap has also been observed in several other quasi-1D
spin-1 materials as
(CH$_3$)$_4$NNi(NO$_2$)$_3$ \cite{gadet91},
AgVP$_2$S$_6$ \cite{mutka91}, and Y$_2$BaNiO$_5$
\cite{y2banio5,xu96} which is probably the best realization
of the spin-1 AF Heisenberg chain.
Finally, it is worth noting that the recent compounds
Ni(C$_5$D$_{14}$N$_2$)$_2$N$_3$(PF$_6$)
(NDMAP) \cite{ndmap}  and Ni(C$_5$H$_{14}$N$_2$)$_2$N$_3$(ClO$_4$)
(NDMAZ) \cite{ndmaz} enable to investigate experimentally
the high magnetic field properties 
of the spin-1 Heisenberg chain \cite{zheludev01}.
From the theoretical point of view,
several methods have been introduced to 
shed light on the nature of the mechanism of the Haldane-gap 
phenomena and to determine the main characteristics of 
this incompressible spin liquid phase.
These approaches are the non-linear sigma model field theory
obtained in the large spin limit \cite{haldane83,affleckrevue},
the valence bond state (VBS) description \cite{aklt},
the Majorana fermions method \cite{tsvelik90}, and 
the restricted Hilbert space approach \cite{gomez89,mikeska92}. 

Probably, the simplest and most appealing approach to 
study the physical properties of the spin-1 AF Heisenberg chain 
consists to add a biquadratic interaction to the model (\ref{heischain}):
\begin{equation}
{\cal H}_{\beta} = J\sum_i \left[{\vec S}_i \cdot {\vec S}_{i+1}
+ \beta \left({\vec S}_i \cdot {\vec S}_{i+1}\right)^2 \right] .
\label{hambiqua}
\end{equation}
One of the main interest of this extended Heisenberg model 
is that for $\beta=1/3$, the so-called AKLT point \cite{aklt},
the model has an exactly solvable ground state which
captures the main characteristics of the Haldane spin liquid phase.
At this special point, the Hamiltonian (\ref{hambiqua}) 
is equivalent to the sum of projection operators 
$P_2(i,i+1)$ that project onto the spin-2 contribution for every 
pair of nearest-neighbor spins:
\begin{equation}
{\cal H}_{AKLT} = {\cal H}_{\beta=1/3}=
2 J\sum_i \left[P_2\left(i,i+1\right) - \frac{1}{3} \right]. 
\label{aklt}
\end{equation}
From this structure, the ground state of this model 
can be constructed exactly using nearest-neighbor valence 
bonds. Following Affleck and coworkers pionneer work 
\cite{aklt}, each original $S=1$ spin is written as 
two spin-1/2 variables in a triplet state.
The ground state is then obtained by coupling into 
a singlet state all nearest-neighbor spin-1/2,
thus forming a crystalline pattern of valence bonds
(see Fig. \ref{akltper}).
\setlength{\unitlength}{1.57cm}
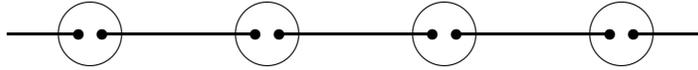
\begin{figure}[h]
\begin{center}
\begin{picture}(5.9,1)
\multiput(0.9,0.4)(1.5,0){4}{\circle*{0.1}}
\multiput(1.1,0.4)(1.5,0){4}{\circle*{0.1}}
\multiput(1,0.4)(1.5,0){4}{\circle{0.5}}
\drawline[10](1.1,0.4)(2.4,0.4)
\drawline[10](2.6,0.4)(3.9,0.4)
\drawline[10](4.1,0.4)(5.4,0.4)
\drawline[10](5.6,0.4)(6.2,0.4)
\drawline[10](0.9,0.4)(0.3,0.4)
\end{picture}
\end{center}                 
\renewcommand{\baselinestretch}{0.8}
\caption{VBS state in a chain with periodic boundary conditions.}
\label{akltper}
\end{figure}
This state is called the valence-bond-solid (VBS) state.
For periodic BC and an even number of sites, 
this VBS state is the unique ground state and does
not break the translation symmetry of the Hamiltonian (\ref{aklt}).
It can also be shown rigorously \cite{aklt} that the AKLT model
has a spectral gap whose value 
has been computed numerically:
$\Delta  \simeq 0.7 J$ \cite{fath93}.
In addition, the authors of Ref. \cite{aklt} have also 
been able to show that the spin-spin correlation function 
decays exponentially with the distance with a finite
correlation length equals to $\xi =1/ \ln 3$ lattice spacings.
The VBS state is therefore the simplest incompressible 
spin liquid phase with a unique ground state and a gap
to all excitations. 

Despite the exponential decay of correlations in 
this spin liquid phase, there is a subtle form of hidden 
AF ordering. 
This hidden order was first
discovered by den Nijs and Rommelse \cite{denijs89} 
using an equivalence of the spin-1 chain to 
a two-dimensional restricted solid-on-solid model
and also later by Tasaki \cite{tasaki91} by means 
of a geometrical approach.
A simple way to exhibit this hidden AF
ordering is to look at the VBS state written in 
the conventional $S^z$ representation.
A typical configuration has the following structure:
$..0 \uparrow 0..0 \downarrow \uparrow \downarrow \uparrow 0 \downarrow 
0..0 \uparrow 0..$, 
i.e. each $\uparrow$ ($S^z_n = +1$ state) 
is followed by $\downarrow$ ($S^z_n = -1$ state)
with an arbitrary number of $0$ states ($S^z_n = 0$ state)
between and vice versa.
If we ignore the spins with $S^z_n = 0$ state
then the remaining spins display a perfect 
long-range spin-1/2 N{\'e}el ordering:
$\uparrow \downarrow \uparrow \downarrow  \uparrow \downarrow ..$.
The VBS state has thus a perfect
dilute AF N{\'e}el order.
Because of the arbitrary number of $S^z_n = 0$ sites
inserted between $S^z_p = \pm 1$ sites, 
the long-range AF order is invisible
in the spin-spin correlation function 
which has an exponential decay. 
However, this hidden order becomes manifest in the 
non-local string order parameter introduced 
by den Nijs and Rommelse \cite{denijs89} and defined by
\begin{equation}
{\cal O}^{\alpha}_{\rm string} = -\lim_{|i-j| \rightarrow + \infty}
\langle S_i^{\alpha} \exp\left(i \pi \sum_{k=j+1}^{i-1} 
S_k^{\alpha} \right)
S_j^{\alpha} \rangle , 
\label{string}
\end{equation}
with $\alpha=x,y,z$. In the VBS ground-state, this order parameter
can be computed exactly \cite{denijs89}: ${\cal O}^{\alpha}_{\rm string} = 4/9$
i.e. this object exhibits long-range order in the VBS phase.
The existence of such a hidden order is essential to 
the basic mechanism of the Haldane gap:
breaking this topological order to create an excitation 
costs a finite energy gap.
The properties of this dilute AF order 
was further elucidated by Kennedy and Tasaki \cite{kennedy92}.
They were able to show, using a non-local unitary transformation,
that the long-range order ${\cal O}^{\alpha}_{\rm string} \ne 0$
and the Haldane gap are related to
a spontaneous breaking of a hidden Z$_2$ $\times$ Z$_2$ symmetry.
In particular, the string order parameter 
becomes the usual local ferromagnetic order parameter under
this non-local unitary transformation.
Another important consequence of this broken Z$_2$ $\times$ Z$_2$ symmetry 
is the existence of a quasi-degeneracy 
of four lowest energy levels for a finite-size chain.
This can also be understood within the VSB state since
for open BC the AKLT model (\ref{aklt})
has exactly four ground states with the existence
of two spin-1/2 degrees of freedom that are 
unpaired at each end of the chain (see Fig. \ref{akltopen}).
\setlength{\unitlength}{1.57cm}
\begin{figure}[h]
\begin{center}
\begin{picture}(5.9,1)
\multiput(0.9,0.4)(1.5,0){4}{\circle*{0.1}}
\multiput(1.1,0.4)(1.5,0){4}{\circle*{0.1}}
\multiput(1,0.4)(1.5,0){4}{\circle{0.5}}
\drawline[10](1.1,0.4)(2.4,0.4)
\drawline[10](2.6,0.4)(3.9,0.4)
\drawline[10](4.1,0.4)(5.4,0.4)
\put(5.6,0.4){\vector(0,1){0.42}}
\put(0.9,0.4){\vector(0,1){0.42}}
\end{picture}
\end{center}
\renewcommand{\baselinestretch}{0.8}
\caption{VBS state with open boundary conditions;
unpaired bonds are left at the boundaries resulting
into two free spin-1/2 objects at the edges.}
\label{akltopen}
\end{figure}
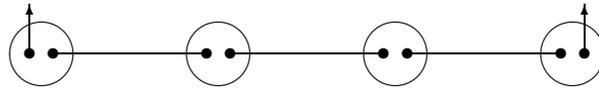                               
Three of these states constitute a spin triplet 
whereas the last one is a spin singlet.

Beside the fact that only the ground state of the Hamiltonian (\ref{aklt})
is known exactly, 
the AKLT approach is also very useful to describe 
low-lying excitations of the model through
a variational approach starting from the VBS state.
A first approach \cite{arovas88}
consists to create a ``magnon'' excitation
by applying a spin-flip on the VBS ground state ($|VBS \rangle$):
\begin{equation}
|k \rangle = \frac{1}{\sqrt{N}} \sum_{j=1}^{N} 
\exp\left(i k j \right) S_j^{+} |VBS \rangle ,
\label{magnonvar}
\end{equation}
$N$ being the system size.
It is very tempting to interpret this excitation as 
a standard $S=1$ magnon i.e. an excitation which modifies 
only locally the AF N{\'e}el order.
In fact, the trial wave function (\ref{magnonvar})
does not describe a magnon excitation 
of the perfect hidden N{\'e}el order of the VBS state
but it has instead a solitonic nature \cite{fath93}.
Indeed, in the  $S^z$ representation,
a typical configuration that appears in 
$S_j^{+} |VBS \rangle$
has the following structure:
$..0 \uparrow 0..0 \downarrow \uparrow 0 \uparrow 0 \downarrow
0..0 \uparrow 0..$ 
Removing the $0$ states, this configuration
becomes $\uparrow \downarrow \uparrow \uparrow \downarrow
\uparrow..$ i.e. a soliton or domain wall which destroys non-locally the 
AF N{\'e}el long-range order.
An alternative approach to describe the 
elementary excitations of the AKLT (\ref{aklt}) model is to start from 
the VBS state and to promote the spins of 
the link $(j,j+1)$  into a triplet state $|\Phi_j^a \rangle$
\cite{knabe88,fath93,mikeska95}.
This low-lying excitation carrying momenta $k$
describes a moving triplet bond, usually called crackion, and it is
defined by
\begin{equation} 
|\Phi^a \left(k\right)\rangle = \frac{1}{\sqrt{N}} \; 
\sum_{j=1}^{N}
\exp\left(i k j \right) |\Phi_j^a \rangle .
\label{tripletmov}
\end{equation} 
In fact, the two approaches (\ref{magnonvar}) and 
(\ref{tripletmov}) are related and describe 
the same dispersion relation for
the triplet excitation \cite{fath93}.
The solitonic nature of the excitation (\ref{tripletmov})
becomes manifest by considering open boundary conditions
and the non-local unitary transformation introduced
by Kennedy and Tasaki \cite{kennedy92}.
In that case, the trial excited wave function corresponds 
to a domain wall that interpolates between two of the 
four possible ground states \cite{fath93}.

One of the main success of this VBS approach
stems from the fact that the properties of the VBS
state are not 
special to the AKLT point (\ref{aklt}) but have a larger extent.
This model turns out to be smoothly connected 
to the spin-1 Heisenberg chain in the sense that
they share the same physical properties.  
In this respect,  numerical investigations 
\cite{kennedy90,biqua,arovas92,hatsugai91,joli96,polizzi98,tsai00}
of the phase diagram of the bilinear-biquadratic 
spin-1 Hamiltonian (\ref{hambiqua}) have revealed that
this model with $|\beta| < 1$ belongs to
a same phase, the Haldane phase,
whose main characteristics are well described
by the VBS state.
In particular, the origin of the Haldane 
gap in the spin-1 Heisenberg chain is 
a consequence of a hidden Z$_2$ $\times$ Z$_2$ 
broken symmetry as for the AKLT model (\ref{aklt}).
At the Heisenberg point $\beta=0$, 
the hidden AF N{\'e}el ordering 
is no longer perfect as at the AKLT point.
It is only locally destroyed so that there is 
still long-range order at $\beta=0$ with a non-zero
string order parameter (\ref{string})
but smaller than at the AKLT point:
${\cal O}^{\alpha}_{\rm string} \simeq 0.374$ 
\cite{arovas92,hatsugai91,dmrg}.
A second consequence of the spontaneous
breaking of the Z$_2$ $\times$ Z$_2$ symmetry
is the two almost free $S=1/2$ degrees of freedom at the ends 
of a finite chain with open BC.
At the Heisenberg point, exact diagonalizations 
\cite{kennedy90} of finite open samples with an even
number of sites have shown that the ground state is 
a singlet and the existence of an exponentially low-lying
triplet state (the so-called Kennedy triplet) in the 
Haldane gap. This leads to 
a fourfold ground-state degeneracy in the thermodynamic 
limit which can be simply understood by means of the VBS 
description of Fig. \ref{akltopen} as described above.
The existence of these two $S=1/2$ edge states
is a robust and striking property of the Haldane spin liquid phase
and  they are present in the whole region $|\beta| < 1$ 
of the bilinear-biquadratic spin-1 model \cite{polizzi98,tsai00}.
The physical properties of these $S=1/2$ chain-boundary
excitations in the open spin-1 chain have been investigated in detail
numerically 
\cite{dmrg,miyashita93,tonegawa93,qin95,legeza97,batista98,polizzi98,tsai00,alet00}
and also analytically \cite{mitra90,sorensen94,phle02}.
Remarkably enough,
this liberation of fractional spin-1/2 
degrees of freedom  in a spin-1 Heisenberg chain
has been observed experimentally in the spin-1 compound 
NENP cut by non-magnetic impurities (Zn$^{2+}$
or Mg$^{2+}$) \cite{glarum91,goto97}
and also doped with magnetic ions (Cu$^{2+}$) \cite{hagiwara90}.
In particular, these spin-1/2 chain-boundary excitations reveal itself 
as satellite peaks in the NMR profile of the Mg-doped 
Y$_2$BaNiO$_5$ \cite{tedoldi99}.
A very recent inelastic neutron scattering experiments \cite{kenzelmann03}
further probe the microscopic structure of these edge states
through the wave-vector dependence of the Zeeman resonance.

\subsection{General spin-$S$ case}

For higher spin-$S$ value ($S > 1$), the difference between 
integer-spin and half-integer AF Heisenberg chains
has been explained within the semiclassical analysis
pionnered by Haldane \cite{haldane83,affleckrevue}
and also by means of the Abelian \cite{schulz86} 
and non-Abelian \cite{affleck87,cabra98} bosonization approaches.
The Haldane's conjecture for $S > 1$ has also been 
investigated numerically and experimentally over the years.

The numerical studies \cite{moreo87,ziman87,hallberg96,lou00}
of the $S=3/2$ AF Heisenberg chain report that the model
diplays a critical behavior that 
belongs to the same universality class as the $S=1/2$
Heisenberg chain. Several quasi-1D AF compounds
with spin 3/2 have been found like 
CsVBr$_3$, CsVCl$_3$ \cite{niel77,itoh95},
and AgCrP$_2$S$_6$ \cite{mutka93}.
In particular, the uniform susceptibility of the spin-3/2
AgCrP$_2$S$_6$ material measured experimentally \cite{mutka93}
is consistent with the result for spin-1/2, i.e. confirming
the universal behavior of half-integer spins.
In the $S=2$ case, the existence of a Haldane gap 
has been established in various quantum Monte-Carlo approaches
\cite{qmcspin2} and in 
DMRG calculations \cite{nishiyama95,jolibis96,qin97}.
In the thermodynamic limit, a finite gap of $\Delta = 0.085(5)J$
and a finite correlation length $\xi = 49 (1)$ lattice
spacings have been found in DMRG \cite{jolibis96}.
Several Haldane spin-2 compounds have been discovered
over the years (see Ref. \cite{yamashita00} for a review)
like CsCrCl$_3$ and (2,2$^{'}$-bipyridine)trichloromanganase(III)
(MnCl$_3$(bipy)) \cite{granroth96}.
Experiments \cite{granroth96} on this last compound 
report the existence of a gap $\Delta/J = 0.07 \pm 0.02$
which is in good agreement with the numerical results.

Integer $S > 1$ spin chains share similar properties
as the $S=1$ case.
In particular, there is still a hidden
AF order in 
the integer $S > 1$ case.
A non-local string order parameter that measures this
topological order has been proposed by Oshikawa \cite{oshikawa92}:
\begin{equation}
{\cal O}^{\alpha}_{\rm string}\left(\frac{\pi}{S}\right)
= -\lim_{|i-j| \rightarrow + \infty}
\langle S_i^{\alpha} \exp\left(i \pi/S \sum_{k=j+1}^{i-1}
S_k^{\alpha} \right)
S_j^{\alpha} \rangle ,
\label{stringS}
\end{equation}
where $\alpha=x,y,z$.
This order parameter can be exactly computed 
in spin-$S$ generalization of the VBS state \cite{oshikawa92,totsuka95}
and numerically for the Heisenberg chain 
\cite{hatsugai92,nishiyama95,jolibis96}.
For instance, in the spin-2 case, 
DMRG calculations \cite{jolibis96} 
found
${\cal O}^{\alpha}_{\rm string}\left(\pi/2\right)
= 0.726(2)$. 
However, the full characterization 
of the hidden AF order for 
integer spin $S>1$  is less clear than for $S=1$
where, as seen above, it is related to a broken 
Z$_2$ $\times$ Z$_2$ symmetry.
In fact, the nature of the hidden symmetry breaking 
is still an open problem for $S>1$.
Insight on this issue might be gained by studying the structure of edge
states of open integer spin-$S$ chains.
Higher-$S$ VBS states with open BC have $(S+1)^2$-fold
degenerate ground states with the presence of 
two free $S/2$ degrees of freedom at each end of the chain in close
parallel to the $S=1$ case (see Fig. \ref{akltopen}).
A non-linear sigma model approach \cite{ng94}
of the open spin-$S$ Heisenberg chain reveals the existence 
of spin-$S/2$ edge states in the integer spin case.
These results lead us to expect that
Z$_{2S}$ $\times$ Z$_{2S}$ might be a good
candidate to describe the hidden symmetry breaking scheme 
in the integer spin-$S$ case. In the $S=2$ case, 
the approximate nine-fold degeneracy of the ground state
of finite open chains have been observed numerically
\cite{qin95,nishiyama95,jolibis96} by DMRG.
The presence of these spin-1 chain-end excitations has
also been confirmed experimentally in the 
ESR study of the spin-2 Haldane compound CsCrCl$_3$
doped with non-magnetic Mg$^{2+}$ ions \cite{yamazaki96}.

The liberation of these fractional spin-$S/2$ degrees
of freedom when non-magnetic impurities 
are introduced in integer spin-$S$ AF Heisenberg
chain is a remarkable fact. 
It stems from the presence of a non-trivial topological
ordering in the ground state of the 
spin-$S$ chain.
In this respect, the existence of these edge states opens 
the possibility to make a distinction between 
spin-$S$ Haldane liquid phases.
These massive spin liquid phases share the 
same thermodynamics properties but have for instance different
chain-end ESR response due to the special nature of their edge states.
In fact, this topological distinction between gapped spin liquid 
phases can also be applied to the massless case.
All half-integer AF Heisenberg spin chains have the 
same bulk universal properties described by an
IR fixed point which belongs to the su(2)$_1$ WZNW universality 
class. However, the ground state of half-integer 
spin chains still sustains a topological ordering. 
Indeed, by means of a non-linear sigma
model approach, Ng \cite{ng94} has given evidence
that edge states with fractionalized quantum 
number $(S-1/2)/2$ exist in
open half-integer spin-$S$ AF Heisenberg chain. 
This interesting result has been confirmed 
numerically by examining 
open spin chains with spins up to $S=9/2$
by DMRG \cite{qin95,lou02}.
Recently, a numerical investigation \cite{qin03} of the non-local
string order parameter (\ref{stringS}) 
directly reveals the presence of a topological order
in half-integer spin-$S$ AF Heisenberg chains.
The existence of this hidden topological order (or the related
chain-end excitations) leads to a topological distinction 
between gapless spin liquid phases with the same emerging
quantum criticality.

\subsection{Two-leg spin ladder}

A further impetus for the study of low-dimensional
spin systems was given by the discovery
of spin-ladder materials one decade ago \cite{dagottorev}.
These systems consist of a finite number ($n_{\rm leg}$)
of spin-1/2 AF Heisenberg chains
coupled by a transverse exchange interaction $J_{\perp}$.
The Hamiltonian of this model reads as follows
\begin{equation}
{\cal H}_{n_{\rm leg}} = J_{\parallel} \sum_i \sum_{a=1}^{n_{\rm leg}}
{\vec S}_{a,i} \cdot {\vec S}_{a,i+1}
+ J_{\perp} \sum_i \sum_{a=1}^{n_{\rm leg} - 1}
{\vec S}_{a,i} \cdot {\vec S}_{a+1,i} ,
\label{laddergen}
\end{equation}   
where ${\vec S}_{a,i}$ is a spin-1/2 operator at 
site $i$ on the a-th chain and the in-chain 
exchange $J_{\parallel}$
is antiferromagnetic ($J_{\parallel} > 0$).
One of the main interest of these spin ladders
stems from the fact that
they are intermediate objects between 1D and 2D systems.
Moreover, these quasi-1D systems share many properties 
with the cuprates and, because of their relative simplicity,
may provide some
insights on mechanisms behind high-T$_c$ superconductivity.

Despite this original motivation to 
study spin ladders,
these systems display striking properties which 
make them interesting by themselves
in the field of 1D quantum magnetism.
In particular,
in close parallel to the qualitative difference 
between integer and half-integer AF Heisenberg spin
chains,
the universal properties of spin ladders (\ref{laddergen})
with open boundary conditions in the transverse
direction strongly
depend on the parity of the number of legs $n_{\rm leg}$
\cite{gopolan94,white94,frischmuth96,rojo96,sierra96}.
Ladders with even number of legs are
spin liquids with a finite gap in the 
excitation spectrum and exponentially decaying 
spin-spin correlations.
The original massless spinons
of the spin-1/2 chains get confined by the interchain
coupling to form coherent optical spin-1 magnon
excitations. Upon doping, these ladders display 
quasi-long-range superconducting pairing correlations with 
an approximate d-wave symmetry \cite{dagottorev}.
In contrast, odd-legged ladders have a gapless 
spectrum which is characterized by a central
charge $c=1$ corresponding to an effective
$S=1/2$ AF Heisenberg spin chain.
Upon doping, odd-legged ladders exhibit a
metallic behavior typical of a Luttinger liquid \cite{revuelutt}.
This difference on the energy spectrum of the spin
ladders depending on the parity of $n_{\rm leg}$
is strongly reminiscent of the Haldane conjecture
for AF Heisenberg spin chains with $S= n_{\rm leg}/2$.
It can be simply understood from the fact that
in the strong ferromagnetic rung coupling limit 
the spin ladder (\ref{laddergen}) is indeed equivalent
to a single $S= n_{\rm leg}/2$ AF Heisenberg chain.

Several spin ladders materials have been 
discovered over the years.
The family of compounds
Sr$_{n-1}$Cu$_{n+1}$O$_{2n}$ approximately realize ladders with 
$n_{\rm leg} = (n+1)/2, n=3,5,7..$ legs \cite{gopolan94}.
The even-odd scenario in ladders was confirmed 
experimentally from
susceptibility and muons resonance  measurements
on this compound with  $n=3$ and $n=5$ \cite{azuma94,kojima95}.
The existence of a spin gap in the two-leg case ($n_{\rm leg}=2$) 
was found in experiments on the two-leg compounds
A$_{14}$Cu$_{24}$O$_{41}$ (A$_{14}$ $\equiv$ La$_6$Ca$_8$,
(Sr,La,Ca)$_{14}$) \cite{carter96},
Cu$_2$(C$_5$H$_{12}$N$_2$)$_2$Cl$_4$ \cite{chaboussant97,hammar98},
(C$_5$H$_{12}$N)$_2$CuBr$_4$ \cite{watson01} 
and the organic compound BIP-BNO \cite{organique}.
However, it is worth noting that additional 
interactions are required to fully describe these materials
such as for instance a non-negligeable ring
exchange for the compound
A$_{14}$Cu$_{24}$O$_{41}$ \cite{brehmer99,nunner} or 
a small diagonal interchain exchange interaction 
for Cu$_2$(C$_5$H$_{12}$N$_2$)$_2$Cl$_4$ \cite{chaboussant97,hammar98}.

The existence of a spin gap in the two-leg case 
can be anticipated by considering a strong-coupling
analysis of the model (\ref{laddergen}) with $n_{\rm leg}=2$. 
In the limit of strong AF interchain
coupling $J_{\perp} \gg J_{\parallel}$, 
the ground state consists of singlet bonds formed
across the rungs of the ladder, with 
triplet excitations separated by a large energy gap
of order $J_{\perp}$. Perturbations in small $J_{\parallel}$
will cause these singlet bonds to resonate and 
the triplet excitations form a band with bandwidth
$\sim J_{\parallel}$ but the spectral gap 
survives \cite{barnes93}.
In the strong ferromagnetic  interchain
$-J_{\perp} \gg J_{\parallel}$ coupling case,            
local spins $S=1$ associated with each rung
of the ladder are formed leading thus 
to a spin-1 AF Heisenberg chain with 
a non-zero Haldane gap in the energy spectrum.
The cross-over between strong and weak coupling
limits has been carefully analysed in numerical
calculations \cite{hida91,watanabe94,nishiyama95bis,white96}.
It turns out that the ground state in the 
two strong-coupling limits $|J_{\perp}|/ J_{\parallel} \gg 1$
evolves adiabatically with increasing $J_{\parallel}$.
The spin gap survives for arbitrarily large
$J_{\parallel}/|J_{\perp}|$ and finally vanishes for $J_{\perp} = 0$,
i. e. when the two spin-1/2 AF Heisenberg chains are decoupled.
The critical point $J_{\perp} = 0$ separates thus 
two strong-coupling massive phases:
a rung-singlet phase for $J_{\perp} > 0$
and a phase with $J_{\perp} < 0$  which is smoothly connected to the 
Haldane phase of the spin-1 chain.

The opening of the spin gap upon switching on 
the interchain coupling can be investigated by means
of the bosonization approach \cite{strong92,shelton96,schulz86}.
In particular, the field theory that accounts for 
the massless spinon-optical magnon transmutation, 
when $J_{\perp}$ is small, corresponds to 
an SO(3) $\times$ Z$_2$ symmetric model of four massive
Majorana (real) fermions \cite{shelton96,nersesyan97}
i.e. four off-critical 2D Ising models 
(for a review see for instance the book \cite{bookboso}).
The resulting low-energy field theory is described
by the following Hamiltonian density \cite{shelton96,nersesyan97}:
\begin{eqnarray}
{\cal H}_{n_{\rm leg}=2} &=& 
-\frac{i v_t}{2}
\; \left({\vec \xi}_R \cdot
\partial_x {\vec \xi}_R -  {\vec \xi}_L \cdot \partial_x {\vec \xi}_L\right) 
- i m_t
\; {\vec \xi}_R \cdot{\vec \xi}_L \nonumber \\ 
&-& \frac{i v_s}{2} 
\; \left(\xi_R^0
\partial_x \xi_R^0 -  \xi_L^0  \partial_x \xi_L^0\right)
-i m_s  \xi_R^0 \xi_L^0 + {\cal H}_{\rm marg},
\label{2-legham}
\end{eqnarray}
where ${\vec \xi}_{R,L}$ are a triplet of right and left-moving
Majorana fermions that describe the $S=1$ low-lying excitations
of the two-leg spin ladder 
and the Majorana fermion $\xi_{R,L}^0$ accounts for 
the singlet excitation.
The masses of these real fermions are given by 
in the weak coupling limit: $m_t= J_{\perp} \lambda^2/2\pi$ and
$m_s = - 3 J_{\perp} \lambda^2/2\pi$, $\lambda$ being 
a non-universal constant. 
These triplet and singlet massive Majorana fermions are weakly coupled by
a marginal perturbation associated to
the last term of Eq. (\ref{2-legham}):
\begin{equation}
{\cal H}_{\rm marg} = \frac{g_1}{2} \left({\vec \xi}_R
\cdot {\vec \xi}_L \right)^2 + 
g_2 \; {\vec \xi}_R \cdot {\vec \xi}_L \; \xi_R^0 \xi_L^0,
\label{margterm}
\end{equation}
with $g_1 = - g_2 = \pi a_0 J_{\perp}/2$, $a_0$ 
being the lattice spacing.
The model (\ref{2-legham}), derived in the 
weak coupling limit $|J_{\perp}| \ll J_{\parallel}$,
is expected to have a larger extent and 
to capture in fact the low-energy properties
of the two-leg spin ladder for arbitrary $J_{\perp}$ 
with a suitable 
redefinition of the masses $m_{s,t}$, velocities $v_{s,t}$, and
coupling constants $g_{1,2}$.
This stems from the continuity between the weak- and
strong-coupling limits in the two-leg spin ladder
observed numerically \cite{hida91,watanabe94,nishiyama95bis,white96}.
In addition, in the strong ferromagnetic 
rung limit $-J_{\perp} \gg J_{\parallel}$,
the singlet excitation described by the Majorana
fermion $\xi_{R,L}^{0}$ are frozen ($|m_s| \to \infty$) so that the low-energy
properties of the model (\ref{2-legham}) are governed by the triplet
magnetic excitations corresponding
to the fermions ${\vec \xi}_{R,L}$.                                
The resulting model coincides with 
the low-energy field theory of the spin-1 AF
Heisenberg chain obtained by Tsvelik \cite{tsvelik90}
by perturbing around  the Babujian-Takhtajan \cite{babu} 
integrable point of the bilinear-biquadratic spin-1 chain (\ref{hambiqua})
with $\beta = -1$. 
The Hamiltonian (\ref{2-legham}) correctly thus captures
the low-energy properties of the two-leg spin ladder
in the limit $-J_{\perp} \gg J_{\parallel}$ 
where it reduces to the spin-1 AF Heisenberg chain.

Despite of its apparent simplicity, the model (\ref{2-legham})  
is not an integrable field theory and takes the form 
a massive SO(3) $\times$ Z$_2$ Gross-Neveu model \cite{grossneveu74}.
However, the leading behavior of the physical quantities
of the two-leg spin ladder can be determined 
by treating the marginal contribution (\ref{margterm})
perturbatively.
As shown in Ref. \cite{shelton96}, this term leads to a
renormalization of the masses and velocities
so that the low-energy description (\ref{2-legham}) 
simply reduces to four independent massive Majorana
fermions, i.e. four decoupled off-critical 2D Ising models.
This mapping onto off-critical 2D Ising models can then 
be exploited to derive the low-energy properties of 
the two-leg spin ladder such as, for instance, 
the spectrum of elementary excitations,
low-T thermodynamics and the leading asymptotics
of spin-spin correlation functions.
To this end, one needs to express the spin operators ${\vec S}_{a,i}$
in terms of the Ising fields.
In the continuum limit, the spin densities separate into
the smooth and staggered parts \cite{affleck87}:
${\vec S}_{a} (x) = {\vec J}_{a L} (x) + {\vec J}_{a R} (x)
+ (-1)^{x/a_0} {\vec n}_{a} (x)$.
The chiral su(2)$_1$ currents, uniform parts of the spin densities, can
be written locally in terms of the Majorana fermions \cite{shelton96,allen97}:
\begin{equation}
{\vec J}_{a R,L}  = - \frac{i}{4} \; {\vec \xi}_{R,L} \wedge 
{\vec \xi}_{R,L} + i \frac{\tau_a}{2} \; {\vec \xi}_{R,L} \xi_{R,L}^{0},
\label{curr2leg}
\end{equation}
with $\tau_1 = +1$ and $\tau_2 = -1$.
In contrast, the staggered magnetizations ${\vec n}_{a}$
are non-local 
in terms of the underlying fermions and express in terms 
of the order ($\sigma_a$) and disorder ($\mu_a$)
of the Ising models \cite{tsvelik90,shelton96,allen97}:
\begin{eqnarray}
{\vec n}_+ &\sim& \left(\mu_0 \mu_1 \sigma_2  \sigma_3,
\mu_0 \sigma_1 \mu_2 \sigma_3,  \mu_0 \sigma_1 \sigma_2 \mu_3 \right) 
\nonumber \\
{\vec n}_- &\sim& \left(\sigma_0 \sigma_1 \mu_2  \mu_3,
\sigma_0 \mu_1 \sigma_2 \mu_3,  \sigma_0 \mu_1 \mu_2 \sigma_3 \right), 
\label{stag2leg}
\end{eqnarray}
where ${\vec n}_{\pm} = {\vec n}_1 \pm {\vec n}_2$. 
Due to the existence of non-zero masses $m_{s,t}$ for 
the fermions in Eq. (\ref{2-legham}), 
all related Ising models are non-critical.
A spectral gap is thus present in the two-leg spin ladder
for all signs of the interchain interaction $J_{\perp}$ and it 
opens linearly with $J_{\perp}$ in the weak coupling
limit. In addition, the Ising description (\ref{curr2leg},\ref{stag2leg})  
of the spin densities allows the calculation of the leading
asymptotics of spin-spin correlations using exact results
of the two-point function of a non-critical Ising model.
Since the signs of the triplet and singlet
masses are always opposite in Eq. (\ref{2-legham}),
it can be shown \cite{shelton96} that
the dynamical spin susceptibility 
displays a sharp single-magnon peak
near $q=\pi/a_0$ and $\omega = |m_t|$ which
reflects the elementary nature of the 
triplet excitations for all signs of $J_{\perp}$.

In this respect, the spin liquid phase of the 
two-leg spin ladder is very similar to the Haldane
phase of the spin-1 chain.
However, it has been stressed recently
that the phase with $J_{\perp} <0$, smoothly connected to
the Haldane phase of the spin-1 chain,
and the rung-singlet phase for $J_{\perp} > 0$
are, in fact, topologically distinct gapped phases \cite{kim00}.
The distinction is intimately related to the short-range
valence bond structure of their ground states.
In close parallel to the classification 
of short-range valence bond configurations on 
a two-dimensional square lattice \cite{read89},
two different topological classes can be defined in 
the one-dimensional case by counting the number 
$Q_y$ of valence bonds crossing an arbitrary vertical line \cite{kim00}.
In the case of the  rung-singlet phase, $Q_y$ is always
even while it is odd for the Haldane phase with $J_{\perp} < 0$.
Two different non-local string order parameters can then be defined
in connection to this topological 
distinction \cite{nishiyama95bis,white96,kim00}:
\begin{eqnarray}
{\cal O}^{\alpha}_{\rm even} &=&-\lim_{|i-j| \rightarrow + \infty}
\langle \left(S_{1,i+1}^{\alpha} + S_{2,i}^{\alpha}\right)
e^{i \pi \sum_{k=j+1}^{i-1}
\left(S_{1,k+1}^{\alpha} + S_{2,k}^{\alpha}\right)} 
\left(S_{1,j+1}^{\alpha} + S_{2,j}^{\alpha}\right) \rangle  \nonumber \\
{\cal O}^{\alpha}_{\rm odd} &=&-\lim_{|i-j| \rightarrow + \infty}
\langle \left(S_{1,i}^{\alpha} + S_{2,i}^{\alpha}\right)
e^{i \pi \sum_{k=j+1}^{i-1}
\left(S_{1,k}^{\alpha} + S_{2,k}^{\alpha}\right)} 
\left(S_{1,j}^{\alpha} + S_{2,j}^{\alpha}\right) \rangle , 
\label{stringorderpar2}
\end{eqnarray}
with $\alpha= x,y,z$.
The Haldane and rung-singlet phases are then characterized by
${\cal O}^{\alpha}_{\rm odd} \ne 0$, ${\cal O}^{\alpha}_{\rm even} =0$,
and by 
${\cal O}^{\alpha}_{\rm even} \ne 0$, ${\cal O}^{\alpha}_{\rm odd} =0$
respectively.
The order parameters (\ref{stringorderpar2}) reveal also 
the difference nature of the hidden AF N{\'e}el order
in the Haldane phase (non-zero triplet states along the 
rung) and in the rung-singlet phase (non-zero triplet states 
along the diagonal) \cite{nishiyama95bis,white96,kim00}.
This topological difference can be discussed in light of
the Ising description (\ref{stag2leg}) of the two-leg spin
ladder. 
The  string order parameters (\ref{stringorderpar2})
can be expressed in terms of the
order and disorder Ising operators \cite{nakamura03}:
${\cal O}^{z}_{\rm odd} \sim \langle \sigma_1 \rangle^2
\langle \sigma_2 \rangle^2$ and 
${\cal O}^{z}_{\rm even} \sim \langle \mu_1 \rangle^2
\langle \mu_2 \rangle^2$.
For a ferromagnetic (respectively antiferromagnetic) 
interchain coupling, the Ising models in the triplet
sector are in their ordered (respectively disordered) phases
so that ${\cal O}^{z}_{\rm odd} \ne 0$ and 
${\cal O}^{z}_{\rm even} = 0$
(respectively ${\cal O}^{z}_{\rm odd} = 0$ and
${\cal O}^{z}_{\rm even} \ne 0$).
In this Ising description, the phases with $J_{\perp} > 0$
and $J_{\perp} < 0$ are thus simply related by a Kramers-Wannier
duality transformation on the underlying Ising models.

The topological distinction between these two 
phases becomes manifest when 
analysing the ground-state degeneracy depending on
the nature of BC used.
In the open BC case, as noted by the authors of Ref. \cite{kim00},
ground states of gapped spin liquid states 
characterized by an odd value of $Q_y$ have 
spin-1/2 edge states, while these end states 
disappear when $Q_y$ is even.
The existence of these $S=1/2$ chain-end degrees of freedom
leads to a ground-state degeneracy in a two-leg
spin ladder with open BC.
In the Haldane phase of the ladder with $J_{\perp} < 0$,
finite open chains with an even number of sites 
have a singlet ground-state with an exponentially
low-lying triplet in the spin gap resulting 
on a fourfold ground-state degeneracy in the thermodynamic
limit. In contrast, 
the ground state  in the rung-singlet phase
is always unique whether open or
periodic BC are used: no low-lying triplet
states are found inside the spin gap
in open ladder with $J_{\perp} > 0$ and an even number of sites.
In this respect, 
the rung-singlet phase with $J_{\perp} > 0$,
in contrast to the $J_{\perp} < 0$ case, is not
equivalent to the Haldane phase characterized by
$S=1/2$ chain-end excitations even though they share
similar properties such as the presence of a spin gap,
and a non-zero string order parameter.
The existence or absence of spin-1/2 edge states
in the open two-leg spin ladder can also be discussed
within the Ising model description (\ref{2-legham}) \cite{phle02}.
In particular, the fourfold ground-state degeneracy 
of the open two-leg ladder with a ferromagnetic interchain
coupling can be obtained within this approach.
Indeed,  for $J_{\perp} < 0$, the
three Ising models for the triplet sector are all in their
ordered phases while the Ising model for the singlet 
degrees of freedom belongs to its disorder phase
so that
$\langle \sigma_i \rangle \ne 0$ ($i=1,2,3$) 
and $\langle \sigma_0 \rangle = 0$.
In that case, each Ising model in the triplet 
sector has a doubly degenerate ground state
which gives thus an eightfold degeneracy.
However, there is a redundancy in the Ising description
since the triplet Hamiltonian in Eq. (\ref{2-legham}),
the total
uniform and staggered magnetizations in Eqs. (\ref{curr2leg},\ref{stag2leg}) are all
invariant under the transformation:
$\xi^{i}_{R,L} \rightarrow - \xi^{i}_{R,L},
\mu_{i} \rightarrow \mu_{i}$, 
and $\sigma_{i} \rightarrow -\sigma_{i}$, $i=1,2,3$.
This leads to a physical fourfold ground-state degeneracy as
it should be in the Haldane phase.                                                             
Another advantage of this Ising model description is
to make explicit the spontaneous breaking of
a hidden Z$_2$ $\times$ Z$_2$ symmetry
associated to the ground-state degeneracy for $J_{\perp} < 0$.
The existence of this hidden symmetry can also be revealed
with help of a non-local unitary transformation  on 
the lattice spins of the two-leg ladder \cite{takada92} as in the spin-1
AF Heisenberg chain \cite{kennedy92}.

Finally, the Haldane and rung-singlet phases of the two-leg spin
ladder can also be distinguished by
their optical properties.
In the rung-singlet phase, 
the existence of singlet or triplet two-magnon bound states below
the two-magnon continuum has been predicted 
theoretically
by a number of groups \cite{damle98,sushkov98,jurecka00,zheng01}.
Recently, the singlet two-magnon bound state has been observed 
in the optical conductivity spectrum of the 
compound (La,Ca)$_{14}$Cu$_{24}$O$_{41}$ \cite{windt01} 
which contains layers with Cu$_2$O$_3$ two-leg spin 
ladder with $J_{\perp} >0$.
On the contrary, it is expected that for a ferromagnetic 
interchain coupling no bound states are present 
below the two-particle continuum as in the spin-1 
AF Heisenberg chain \cite{dmrg}.
The existence of bound states in the two-leg 
spin ladder can be discussed in the context of the field 
theory approach (\ref{2-legham}).
Indeed, the marginal term (\ref{margterm})
plays its trick by giving rise to an effective interaction between 
the magnon excitations.
In the AF interchain case, the contribution (\ref{margterm}) 
is in fact marginal relevant and the interaction between
magnons are attractive.
The presence of bound states in this case
can be argued qualitatively
as a result of this attractive interaction.
The Hamiltonian (\ref{margterm})
takes the form of a SO(4) Gross-Neveu model up to 
irrelevant contributions and a duality transformation
$\xi^0_R \rightarrow - \xi^0_R$ in the singlet sector.
The latter model is integrable \cite{zamolo79} and, for
$J_{\perp} >0$, the low-lying excitations are 
massive kinks and anti-kinks interpolating between
the two degenerate ground states resulting from
the spontaneous symmetry breaking of the Z$_2$ symmetry: 
$\xi_R^a \xi_L^a \rightarrow - \xi_R^a \xi_L^a$, $a=0,..,3$.
The mass terms in the full 
Hamiltonian (\ref{2-legham}) break explicitely 
this Z$_2$ symmetry and the ground-state degeneracy 
is lifted so that kink configurations are no 
longer asymptotic states of the field theory.
The situation is in close parallel to the 
two-dimensional Ising model in its low-temperature phase upon 
switching on a magnetic field.
The mass terms are then expected to induce a linear 
confining potential between the kinks giving rise
to a sequence of bound states.
In contrast, when $J_{\perp} <0$, the perturbation 
(\ref{margterm}) is a marginal irrelevant contribution 
and the spectrum of the model (\ref{2-legham})
consists of massive fermions and their multiparticle excitations.

\subsection{Non-Haldane spin liquid}

The Haldane and rung-singlet phases of the two-leg spin 
ladder are not the only possible gapped spin-liquid states
available in 1D unfrustrated quantum magnets.
In this respect, Nersesyan and Tsvelik \cite{nersesyan97}
have discussed the example of a gapped spin liquid phase
without any coherent magnon excitations.
The spectral function of this state displays a broad feature
rather than a single sharp magnon peak.
Such a spin liquid can be stabilized by 
the introduction of a four-spin interchain interaction
that couples two spin-1/2 AF Heisenberg chains:
\begin{equation}
{\cal H}_{\rm so}  =
J \sum_i \sum_{a=1}^{2}{\vec S}_{a,i} \cdot {\vec S}_{a,i+1}
+ K \sum_i \left({\vec S}_{1,i} \cdot {\vec S}_{1,i+1}\right)
\left({\vec S}_{2,i} \cdot {\vec S}_{2,i+1}\right).
\label{hamso}
\end{equation}              
The biquadratic interaction represents
an interchain coupling for
the spin-dimerization operators ($\epsilon_{a,i} \sim (-1)^{i}
{\vec S}_{a,i+1} \cdot {\vec S}_{a,i}$) of each chain 
which can be effectively generated by spin-phonon interaction.
A second motivation to investigate the effect 
of this four-spin interaction stems from orbital degeneracy.
In most of transition metal compounds, in addition to
the usual spin degeneracy, the low-lying electron states 
are also characterized by orbital degeneracy \cite{kugel82}.
A  starting point to study magnetic properties
of magnetic insulators with Jahn-Teller ions is
the two-band Hubbard-like models. 
At quater-filling (one electron
per atom), this system is a Mott insulator in the
limit of strong Coulomb interaction and the state of each
ion can be characterized by a spin degrees of freedom 
${\vec S}_{1,i}$ and an orbital state described 
by a pseudo-spin-1/2 ${\vec S}_{2,i}$.
In the large Coulomb repulsion limit,
the simplest Hamiltonian
that describes the competition between spin and orbital
degrees of freedom in one dimension reduces to
the spin-orbital model (\ref{hamso}).
In particular, this model has been introduced by Pati et al.
\cite{pati98} to explain the unsual magnetic propeties
of the quasi-one-dimensional spin gapped 
material Na$_2$Ti$_2$Sb$_2$O \cite{axtell97}.

The spin-orbital model (\ref{hamso}) is 
unfrustrated for $K <0$ whereas
the frustration manifests itself in the antiferromagnetic
case $K > 0$ only in the intermediate
regime $K \simeq J$ as we shall see later.
The Hamiltonian (\ref{hamso}) is invariant under independent
SU(2) rotations in the spin (${\vec S}_{1}$) and
orbital (${\vec S}_{2}$) spaces. For generic couplings,
the model (\ref{hamso}) is thus SU(2) $\times$ SU(2) $\times$
Z$_2$ symmetric, the additional Ising symmetry being the
exchange between the spins ${\vec S}_{1}$ and ${\vec S}_{2}$.
In the weak coupling limit $|K| \ll J$,
this underlying SO(4) symmetry 
of the spin-orbital model (\ref{hamso}) 
is reflected in the form of its low-energy
Hamiltonian density 
which is described in terms of four 
massive Majorana fermions \cite{nersesyan97,orignac00}:
\begin{eqnarray}
{\cal H}_{\rm so} \simeq 
-\frac{i v}{2}\sum_{a=0}^{3}
\; \left(\xi_R^{a} 
\partial_x \xi_R^{a} 
-  \xi_L^{a} \partial_x \xi_L^{a}\right)
- i m \sum_{a=0}^{3} \; \xi_R^{a} \xi_L^{a} ,
\label{hamso4}
\end{eqnarray}
with $m= \alpha K$, $\alpha$ being a positive non-universal
constant.
In Eq. (\ref{hamso4}),
a marginal contribution, similar to Eq. (\ref{margterm}),
has been neglected.
In contrast to the low-energy description (\ref{2-legham}) 
of the standard two-leg spin ladder,
the triplet and singlet Majorana modes become equally important here
as a consequence of the SO(4) symmetry of the 
spin-orbital model (\ref{hamso}).
The resulting low-energy 
properties of the model 
can then be determined by this
mapping onto four off-critical 2D Ising models.
As the staggered magnetizations (\ref{stag2leg}),
the spin dimerization fields $\epsilon_{1,2}$
can be expressed in terms of the order 
and disorder operators of the underlying Ising models:
\begin{equation}
\epsilon_a \sim \mu_1 \mu_2 \mu_3 \mu_4 
\pm \tau_a \sigma_1 \sigma_2 \sigma_3 \sigma_4 . 
\label{dimerfields}
\end{equation}
For an AF biquadratic coupling ($K >0$), 
the four Ising models belong to their ordered phases
so that the model (\ref{hamso4}) enters a spontaneously dimerized phase 
with a finite gap and
$\langle \epsilon_1 \rangle = - \langle \epsilon_2 \rangle = \pm |\epsilon_0|$.
The ground state is thus two-fold degenerate 
and dimerizes with an alternating pattern 
as shown in Fig. \ref{dimerpi} (a).
\setlength{\unitlength}{1.37cm}
\begin{figure}[ht]
\begin{center}
\begin{picture}(6,5)
\multiput(0.1,0.4)(1,0){6}{\circle*{0.1}}
\multiput(0.1,1.4)(1,0){6}{\circle*{0.1}}
\multiput(0.1,3.4)(1,0){6}{\circle*{0.1}}
\multiput(0.1,4.4)(1,0){6}{\circle*{0.1}}
\put(0.6,0.4){\oval(1.1,0.14)}
\put(2.6,0.4){\oval(1.1,0.14)}
\put(4.6,0.4){\oval(1.1,0.14)}
\put(0.6,1.4){\oval(1.1,0.14)}
\put(2.6,1.4){\oval(1.1,0.14)}
\put(4.6,1.4){\oval(1.1,0.14)}
\put(0.6,4.4){\oval(1.1,0.16)}
\put(2.6,4.4){\oval(1.1,0.16)}
\put(4.6,4.4){\oval(1.1,0.16)}
\put(1.6,3.4){\oval(1.1,0.16)}
\put(3.6,3.4){\oval(1.1,0.16)}
\put(5.6,3.4){\oval(1.1,0.16)[l]}
\put(-0.4,3.4){\oval(1.1,0.16)[r]}
\put(-0.5,5.0){\makebox(0,0){\bf (a)}}
\put(-0.5,2.3){\makebox(0,0){\bf (b)}}
\end{picture}
\end{center}
\renewcommand{\baselinestretch}{0.8}
\caption{
Ground states of the spin-orbital model in
the weak coupling limit $|K| \ll J$. The bonds indicate 
a singlet pairing between the spins.
(a) Alternating dimerization for $K>0$;
(b) in-phase dimerization for $K<0$.
In each case, a second ground state is obtained 
by applying the one-step translation symmetry on each chain
to the states depicted here.}
\label{dimerpi}
\end{figure}
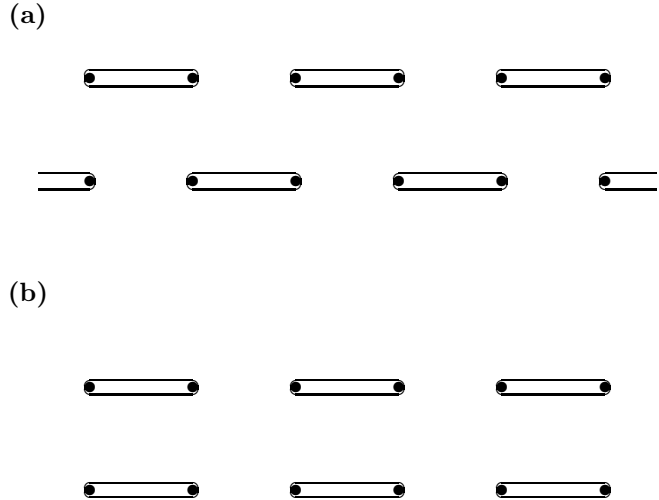                           
In contrast, in the ferromagnetic $K < 0$ case,
the dimerization is now in-phase between the
two chains: 
$\langle \epsilon_1 \rangle = \langle \epsilon_2 \rangle = \pm |\epsilon_0|$
as depicted by Fig. \ref{dimerpi} (b). 
A first distinction between 
this gapped spin liquid state
and the Haldane and rung-singlet phases of 
the two-leg spin ladder stems from the ground-state degeneracy
and the fact that the lattice translation symmetry is 
spontaneously broken in the ground states of Fig. \ref{dimerpi}. 
A more drastic difference appears at the level of
the low-lying excitations.
\setlength{\unitlength}{1.57cm}
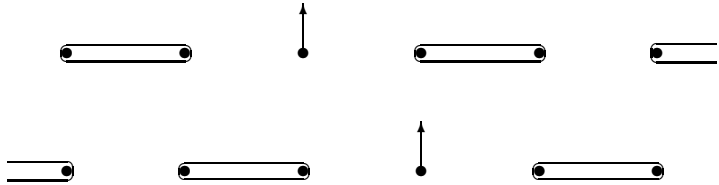
\begin{figure}
\begin{center}
\begin{picture}(5.9,1.8)
\multiput(0.1,0.4)(1,0){6}{\circle*{0.1}}
\multiput(0.1,1.4)(1,0){6}{\circle*{0.1}}
\put(1.6,0.4){\oval(1.1,0.14)}
\put(4.6,0.4){\oval(1.1,0.14)}
\put(0.6,1.4){\oval(1.1,0.14)}
\put(3.6,1.4){\oval(1.1,0.14)}
\put(5.6,1.4){\oval(1.1,0.16)[l]}
\put(-0.4,0.4){\oval(1.1,0.16)[r]}
\put(3.1,0.4){\vector(0,1){0.42}}
\put(2.1,1.4){\vector(0,1){0.42}}
\end{picture}
\end{center}
\renewcommand{\baselinestretch}{0.8}
\caption{Triplet excitation of the alternating
dimerization described in terms of a pair of 
dimerization kinks of each chain.}
\label{dimerpiex}
\end{figure}                              
Indeed, the elementary excitations of the spin-orbital model
are neither optical magnons nor massive spinons but a pair
of propagating massive triplet or singlet kinks connecting
two spontaneously dimerized ground states \cite{nersesyan97}.
An example of such a triplet excitation is described
in Fig. \ref{dimerpiex} for the 
staggered dimerization phase with $K > 0$.
The composite nature of the $S=1$ excitation 
reveals itself in the dynamical structure factor
of the model which can be determined using 
the Ising description (\ref{stag2leg}) of the 
total and relative staggered magnetizations \cite{nersesyan97}.
The dynamical magnetic susceptibility
displays a two-particle threshold instead of 
a sharp magnon peak near $q=\pi/a_0$
as in the two-leg spin ladder or the spin-1 chain.
This incoherent background in the dynamical 
structure factor leads to a new gapped spin liquid 
phase in unfrustrated quantum magnetism
without any coherent magnon excitations.
In this respect, 
this state has been called 
non-Haldane spin liquid by
Nersesyan and Tsvelik \cite{nersesyan97}.
The existence of this spin liquid phase has been
confirmed non-perturbatively at two special points
in the phase diagram of the spin-orbital model (\ref{hamso}).
At $J=3K/4$, the ground state is exactly known \cite{kolezhuk98}
and is a product of checkerboard-ordered spin and orbital
singlets as in Fig. \ref{dimerpi} (a) obtained in the weak
coupling limit.
At $K = -4 J$, the model is exactly solvable
and the ground-state energy and triplet energy gap have been
determined exactly \cite{martins00}. The system has two
spontaneously dimerized ground states and belongs to the
class of non-Haldane spin liquid.

Finally, it is worth noting that the 
alternating and in-phase dimerization phases can 
be distinguished in close parallel to the topological 
distinction between the two gapped phases of the two-leg spin ladder. 
Using the topological criterion of Kim et al. \cite{kim00},
one observes from Fig. \ref{dimerpi} that 
the number $Q_y$ of valence bonds crossing an vertical line
is odd (respectively even) for the staggered (respectively
in-phase) dimerization. The two non-Haldane spin liquid phases 
with $K > 0$ and $K < 0$ 
belong thus to two different topological classes.
This difference becomes manifest when investigating the 
structure of the edge states of the two phases
with open BC.
The physical properties of these boundary excitations 
of the semi-infinite spin-orbital model (\ref{hamso}) can
be determined by means of the Ising mapping (\ref{hamso4}) 
of the weak coupling limit similarily to the cut
two-leg ladder \cite{phle02}.
For $K < 0$, such edge states are absent whereas,
in the  $K > 0$ case, two spin-1/2 chain-end excitations 
are expected \cite{preprint03}.

\section{Frustration effects}     

\begin{figure}[ht]
\begin{center}
\noindent
\epsfxsize=0.5\textwidth
\epsfbox{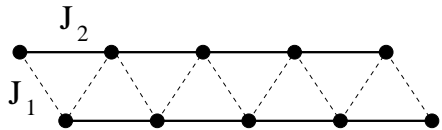}
\end{center}
\caption{\label{zigagfig}%
Two-leg zigzag ladder.
}
\end{figure}

In this section, we shall review some of the main aspects
of the interplay between
frustration and quantum fluctuations in
AF spin chains and spin ladders.
The natural question is whether frustration can stabilize new types of
spin liquid phases with exotic
spin excitations not encountered in Section 2.
The paradigmatic model to analyse the effect of frustration 
in spin chains is the $J_1-J_2$ spin-1/2 Heisenberg chain 
with Hamiltonian:
\begin{equation}
{\cal H} =
J_1 \sum_i {\vec S}_i \cdot {\vec S}_{i+1}
+ J_2 \sum_i {\vec S}_i \cdot {\vec S}_{i+2} ,
\label{j1j2ham}
\end{equation}  
where the next-nearest neighbor coupling $J_2 > 0$ 
is a competing AF interaction which introduces 
frustration.
This model can also be viewed 
as a frustrated two-leg spin ladder where the spin chains 
are coupled in a zigzag way as shown in Fig. \ref{zigagfig}.
Let us first discuss some of the main characteristics of frustration 
in 1D spin systems by means of a semiclassical approach.

\subsection{Semiclassical analysis}
Classically, one of the main effect of frustration is to favor 
non-collinear magnetic ordering where the spins lie
in a plane rather than along a single direction
as in unfrustrated spin systems:
\begin{equation}
{\vec S}_i = S \cos\left( {\vec Q} \cdot {\vec r}_i \right) 
{\vec n}_1 +  S \sin\left({\vec Q} \cdot {\vec r}_i \right) 
{\vec n}_2, 
\label{noncolinear}
\end{equation}
where ${\vec r}_i$ is the spatial location of the site $i$ and
${\vec n}_{1,2}$ are two mutually orthogonal fixed vectors in spin space with
unit length. The magnetic structure corresponding to the ordering (\ref{noncolinear})
is a circular spiral with a pitch angle related to the wave-vector ${\vec Q}$
which is, in general, incommensurate.
In the case of the $J_1-J_2$ chain (\ref{j1j2ham}),
the spiral ground state is stabilized when 
$J_2/J_1 > 1/4$. The spins are arranged in a canted 
configuration in which each spin makes an angle $\alpha$
with its predecessor such that $\cos \alpha = - J_1/4J_2$.
The classical ground state of the model is doubly degenerate 
since the spin configurations (\ref{noncolinear}) 
can turn clockwise and counterclockwise
with the same energy along the
spiral axis ${\vec n}_3 = {\vec n}_1 \wedge {\vec n}_2$.
A Z$_2$ discrete symmetry characterized by 
this right- and left-handed chirality is thus spontaneously broken
in this helical structure. 
A corresponding chiral ordering \cite{villain77} can 
be defined and detected by the chiral order parameter:
$\langle ({\vec S}_i \wedge {\vec S}_{i+1}) \cdot {\vec n}_3 \rangle \ne 0$.
In contrast to the N{\'e}el state of unfrustrated magnets, 
the SU(2) spin rotation of the Heisenberg Hamiltonian is 
completely broken in the non-collinear state  (\ref{noncolinear}).
Three Goldstone or spin-wave modes are thus expected here  from this 
spontaneous symmetry breaking scheme. It leads to a CFT with central 
charge $c=3$
with extended criticality in comparison to the unfrustrated case
($c=2$, see Section 2).
A second distinction stems from the spin-wave calculation \cite{jolicoeur89}
for non-collinear ordered states which reports the presence of two different
spin-wave velocities.
Therefore, the low-energy field theory describing
canted magnets is expected to be non-Lorentz invariant.
To identify this effective field theory,
one needs to define a suitable order parameter
for canted magnets.
The long-wavelength fluctuations of the spin
configurations (\ref{noncolinear}) can be captured
by introducing
a two-component complex field $z_a, a=1,2$ 
of unit modulus ($|z_1|^2 + |z_2|^2 =1$) such that \cite{sachdev02}:
\begin{equation}
{\vec n}_1 + i {\vec n}_2 = \epsilon_{ac} z_c {\vec \sigma}_{a b} z_b,
\label{paraflu}
\end{equation}
where $\epsilon_{ab}$ is the standard antisymmetric tensor
and ${\vec \sigma}$ is a vector formed by the Pauli matrices.
It is indeed straightforward to see that this identification (\ref{paraflu})
correctly reproduces the constraints: ${\vec n}_1^2 = {\vec n}_2^2 =1$
and ${\vec n}_1 \cdot {\vec n}_2 =0$
if the two-component complex field $z_a$ 
belongs to the three-dimensional surface of the unit
sphere in four dimensions denoted by S$^3$.
Moreover, it is worth noting that the field
$z=(z_1,z_2)$ transforms like a $S=1/2$
spinor under spin rotations.
The representation (\ref{paraflu}) is double-valued since
$z$ and $-z$ describe the same non-collinearly ordered state.
The order parameter space corresponding to 
canted magnets (\ref{noncolinear}) is thus  S$^3/Z_2$.

A crucial consequence of this identification
is that the order parameter allows
topologically nontrivial 
vortices having a $2 \pi$ circulation 
since $\pi_1( S^3/Z_2) = Z_2$ \cite{mermin79}.
Upon encircling such a Z$_2$ vortex, called now 
a vison \cite{senthil00}, by a closed loop,
the value of the spinor complex field $z=(z_1,z_2)$
changes smoothly from $z$ to $-z$, i.e. two points diametrally
opposed on the surface of a sphere S$^3$.
The existence of these stable vison defects reveals
itself in the semiclassical description of frustrated 
spin chains like the $J_1-J_2$ Heisenberg chain (\ref{j1j2ham}).
In particular, in sharp contrast to unfrustrated spin
chains, no Pontryagin topological term has been 
found in the effective action that describes
the low-energy properties of the $J_1-J_2$ Heisenberg chain
\cite{rao94,allen95} and two-dimensional frustrated 
Heisenberg antiferromagnets \cite{dombre89}
in the large spin limit.
This stems from the very special topological nature 
of the order parameter for canted magnets 
which has a trivial second homotopy group:
$\pi_2( S^3/Z_2) = 0$ \cite{mermin79}.
However, there is still a subtle distinction between
half-integer and integer spins in the $J_1-J_2$
Heisenberg chain \cite{haldane00,rao94}
related to the existence of stable visons.
A Berry phase calculation \cite{rao94,haldane00,henley92,loss92}
shows that tunneling between sectors with
different Z$_2$ topogical number
is possible for integer spins but not 
for half-integer spins due to cancellation 
between pairs of paths.
The ground state of the $J_1-J_2$ model
is thus non-degenerate for integer spins
and two-fold degenerate 
(two different sectors non-coupled by tunneling effect)
for half-integer spins.
The quantum number carried by elementary excitations 
depends also strongly on the nature of the spins \cite{haldane00}.
In the half-integer case, 
there is an energy gap towards the creation 
of Z$_2$ visons so that these topological defects 
are strongly suppressed.
The low-energy excitations are then described
by the spinor field $z=(z_1,z_2)$ of Eq. (\ref{paraflu})
which is free to propagate in absence of visons
and carries a $S=1/2$ quantum number under 
spin rotations. Neglecting anisotropy, 
the Euclidean effective action that governs 
the leading low-energy properties of the $J_1-J_2$ chain
for half-integer spins is the SU(2) $\times$ SU(2) 
principal chiral model defined by
\begin{equation}
{\cal S}_G = \frac{v}{2g} \int dx d \tau \; 
{\rm Tr} \left(\partial_{\mu} G^{\dagger}
\partial_{\mu} G \right),
\label{chiralprincipal}
\end{equation}
where $G$ is a SU(2) matrix formed by the 
two complex fields $z_{1,2}$.
This field theory \cite{polyakov83,wiegmann87}
is integrable and its low-energy spectrum consists
of massive excitation with $S=1/2$ quantum number
i.e. massive deconfined spinons.
In contrast, in the integer spin case,
the visons now proliferate in the ground state
so that the spinor field $z=(z_1,z_2)$ cannot any longer
be defined as single-valued configurations.
The suitable fields that describe
the low-energy excitations of the 
$J_1-J_2$ chain with integer spins are bilinear of $z$
and correspond to the ${\vec n}_{1,2}$ fields of Eq. (\ref{paraflu}).
In that case, neglecting anisotropy, the effective field theory 
that captures the leading low-energy properties of
the frustrated spin chain 
with integer spins is the non-linear sigma model SO(3) $\times$ SO(3)
with action:
\begin{equation}
{\cal S}_R = \frac{v}{2g} \int dx d \tau \; 
{\rm Tr} \left(\partial_{\mu} R^{-1}
\partial_{\mu} R \right),
\label{chiralprincipalrot}
\end{equation}
where $R$ is a rotation matrix made of the triplet 
vectors ${\vec n}_{1,2}$ and ${\vec n}_{3} = {\vec n}_{1} 
\wedge {\vec n}_{2}$.
The low-lying spectrum of this field 
theory can be determined by a large N approach \cite{rao94,allen95}
or by means of a strong-coupling analysis \cite{haldane00}: 
it consists of a triplet of $S=1$ massive magnons.
In the integer spin case, the spinons are thus expected to be confined 
into optical magnons.

An important prediction, obtained in this semiclassical 
description, is thus the existence of a gapped 
spin liquid phase for half-integer spins with a two-fold degenerate
ground state and massive deconfined spinons.
Such a state is stabilized by frustration
and represents a spin liquid phase not encountered
in Section 2.
We shall now consider the ultra-quantum case with $S=1/2$
where a combination of numerical and field theoretical
techniques can be used to fully determine the main 
characteristics of this spin liquid phase.

\subsection{Spin liquid phase with massive deconfined spinons}

The phase diagram of the spin-1/2 $J_1-J_2$ Heisenberg
chain (\ref{j1j2ham}) has been studied extensively
over the years after the bosonization 
analysis of Haldane \cite{haldane82}.
This problem is not also a purely academic question
since inorganic compounds such as CuGeO$_3$ \cite{hase93,boucher96}
or LiV$_2$O$_5$ \cite{isobe96} can be considered as prototypes
of the  spin-1/2 $J_1-J_2$ chain (\ref{j1j2ham}).
In particular, values such as $J_1 \simeq 160$ K and
$J_2/J_1 \simeq 0.36$ have been proposed
for CuGeO$_3$ \cite{castilla95}.
In addition, the quasi-1D compound SrCuO$_2$ 
contains a collection of spin-1/2 Heisenberg chains
assembled pairwise in an array of weakly 
coupled zigzag ladders \cite{matsuda95,motoyama96}.
These zigzag ladders are built from corner-sharing
Cu-O chains with an exchange $J_2$
staked pairwise in edge-sharing geometry.
The frustrating interaction $J_1$ between  the chains
(see Fig. \ref{zigagfig}) stems from the nearly 
$90^{o}$ Cu-O-Cu bonds and is 
expected to be weak \cite{matsuda95}.
Finally, two other possible realizations 
of the spin-1/2 $J_1-J_2$ chain (\ref{j1j2ham})
have been recently proposed:
the compound (N$_2$H$_5$)CuCl$_3$ which can be 
described as a two-leg zigzag spin ladder with
$J_1/J_2 \simeq 0.25$ \cite{maeshima02}
and Cu[2-(2-{\rm aminomethyl}){\rm pyridine}]Br$_2$
with a ratio $J_2/J_1 \simeq 0.2$ \cite{kikuchi00}.                               

A starting point for investigating the phase diagram 
of the spin-1/2 $J_1-J_2$ chain (\ref{j1j2ham})
is to consider the weak coupling limit when $J_2 \ll J_1$.
This enables us to study the stability 
of the massless spinons of the spin-1/2 AF Heisenberg chain
upon switching on a small next-nearest-neighbor frustrating
interaction $J_2$. 
In this weak-coupling regime, the low-energy effective Hamiltonian 
density of the model reads as follows \cite{haldane82}:
\begin{equation}
{\cal H}_{\rm eff} = \frac{2\pi v}{3} \left( {\vec J}_L^2 
+ {\vec J}_R^2 \right)+ \gamma {\vec J}_L \cdot {\vec J}_R ,
\label{curcurham}
\end{equation}                
where ${\vec J}_{L,R}$ are the left and right su(2)$_1$
currents that generate the su(2)$_1$ quantum criticality 
of the spin-1/2 AF Heisenberg chain.
The low-energy physics of the model are thus mainly
determined by the 
marginal current-current interaction of Eq. (\ref{curcurham})
with coupling constant 
$\gamma \simeq J_2 - J_{2c}$,  $J_{2c}$ being a non-universal
positive constant.
For a small value of $J_2$, one has $\gamma < 0$ 
so that the interaction in Eq. (\ref{curcurham})
is a marginal irrelevant contribution.
The low-energy physics of the $J_1-J_2$ chain 
is thus identical to that of the spin-1/2 AF Heisenberg
chain which is governed by the su(2)$_1$ fixed point
with additional logarithmic corrections introduced
by the current-current interaction.
In this respect, 
frustration plays no important role in the regime $J_2 < J_{2c}$
and the spin-1/2 Heisenberg phase with massless spinons
is stable upon switching on a small value of $J_2$.
However, for $J_2 > J_{2c}$ ($\gamma > 0$), 
the current-current interaction becomes marginal 
relevant and a strong coupling regime develops 
with a dynamically generated spectral gap.
A phase transition of Berezinskii-Kosterlitz-Thouless (BKT) type \cite{bkt} 
occurs at $J_2 = J_{2c}$ which separates 
the gapless spin-1/2 Heisenberg phase from
a fully massive region.
The actual value of the transition
has been determined numerically by different
groups to be: $J_{2c} \simeq 0.2411 J_1$ 
\cite{okamoto92,chitra95,eggert96,whitebis96}.
At this point, it is worth noting that 
the quasi-1D material Cu[2-(2-{\rm aminomethyl}){\rm pyridine}]Br$_2$
can be described by a spin-1/2 $J_1-J_2$ Heisenberg chain with 
$J_2/J_1 \simeq 0.2$ \cite{kikuchi00}
and should thus belong to the critical Heisenberg phase with $J_2 < J_{2c}$.
The absence of a spin gap for this compound 
has been reported experimentally \cite{kikuchi00}.

The main characteristics of the 
strong coupling massive phase with $J_2 > J_{2c}$
can be determined from Eq. (\ref{curcurham}).
The field theory (\ref{curcurham})
is indeed integrable and corresponds to a chiral Gross-Neveu
model or a non-Abelian version of the Thirring model
with a SU(2) symmetry.
The low-energy excitations of the model (\ref{curcurham})
are a massive doublet (massive spinons) with mass 
$m \sim a_0^{-1}
\exp ( - 2 \pi/\gamma)$ \cite{andrei79,belavin79}.
A striking effect of frustration is thus the formation
of a mass for the spinons of the spin-1/2 AF Heisenberg
chain without confining them into $S=1$ excitations
as in the two-leg spin ladder.
A second consequence of frustration is the 
presence of spontaneously dimerization in the model.
A simple way to exhibit this dimerization 
is to use the bosonization approach
to express the Hamiltonian (\ref{curcurham})
in terms of a  $\beta^2 = 8 \pi$ 
sine-Gordon model \cite{haldane82}:
\begin{equation}
{\cal H}_{\rm eff} = \frac{v}{2} \left[ \left(\partial_x \Phi \right)^2
+ \left(\partial_x \Theta \right)^2  \right]
+ \frac{\gamma}{2\pi} \partial_x \Phi_R \partial_x \Phi_L
- \frac{\gamma a_0^2}{4\pi^2}
\cos \sqrt{8 \pi} \Phi ,
\label{sinegordon8pi}
\end{equation}
where $\Phi_{R,L}$ are the chiral components of the bosonic
field $\Phi$ ($\Phi = \Phi_{R} + \Phi_{L}$)  
and $\Theta$ is its dual field ($\Theta = \Phi_{L} - \Phi_{R}$). 
The SU(2) invariance of the model (\ref{sinegordon8pi}) is
hidden in the structure of the interaction 
with a single coupling constant and the 
fact that the bosonic field $\Phi$
is compactified on a circle with a special
radius $R = 1/\sqrt{2\pi}$ consistent with the SU(2) symmetry.
This compactification leads to the following identification:
\begin{equation} 
\Phi \sim \Phi + 2 \pi R n = \Phi + n\sqrt{2 \pi} ,
\label{compact}
\end{equation}       
$n$ being integer. 
In the phase with $J_2 > J_{2c}$,
one has  $\gamma > 0$ so that the bosonic field
is pinned at one of its minima:
$\langle \Phi \rangle = p \sqrt{\pi/2}$, $p$ being integer.
However, by taking into account the identification (\ref{compact}), 
the $\beta^2 = 8 \pi$ sine-Gordon model (\ref{sinegordon8pi})
has only two inequivalent ground states with 
$\langle \Phi \rangle = 0$
and $\langle \Phi \rangle = \sqrt{\pi/2}$.
This two-fold degeneracy can be interpreted as resulting
from the spontaneous breaking of a discrete Z$_2$ symmetry.
This symmetry identifies with 
the one-step translation symmetry (${\cal T}_{a_0}$)
which is described by the following shift on the 
bosonic field \cite{affleckhouches}:
$\Phi \rightarrow \Phi + \sqrt{\pi/2}$.
This symmetry is spontaneously broken in the
phase $J_2 > J_{2c}$ and the two ground-state 
field configurations are connected by 
this translation symmetry.
An order parameter designed to characterize this phase
is the spin dimerization operator
$\epsilon_n = (-1)^n {\vec S}_n \cdot {\vec S}_{n+1}$
which admits the following bosonic representation
in the continuum limit 
$\epsilon \sim \cos \sqrt{2 \pi} \Phi$ \cite{affleckhouches,bookboso}.
This operator changes sign under the lattice translation symmetry
and has a non-zero 
expectation value $\langle \epsilon \rangle \ne 0$ 
in the two ground states of the 
$\beta^2 = 8 \pi$ sine-Gordon model (\ref{sinegordon8pi}).

For $J_2 > J_{2c}$, frustration stabilizes thus a 
gapful spontaneously dimerized phase 
which is characterized by a two-fold degenerate 
ground state and a spontaneous breaking 
of the lattice translation symmetry.
The elementary excitations of this phase
are massive spinons that carry $S=1/2$ quantum number
and identify with the kinks of the underlying dimerization.
A simple way to understand the emergence of 
these deconfined spinons is 
to consider the Majumdar-Ghosh (MG)  \cite{majumdar69} 
point at  $J_2 = J_1/2$
where the ground state of the lattice model (\ref{j1j2ham})
is exactly known.
This MG point for the spontaneously dimerized phase plays a similar
role than the AKLT point (\ref{aklt}) 
for describing the main properties of 
the Haldane phase in the phase diagram of the spin-1 bilinear-biquadratic
chain (\ref{hambiqua}).
For $J_2 = J_1/2$, the Hamiltonian (\ref{j1j2ham}) 
takes the following form up to a constant:
\begin{equation}
{\cal H}_{\rm MG} = \frac{3 J_1}{4} \sum_i P_{3/2}\left(i-1,i,i+1\right),
\label{MG}
\end{equation}
where $P_{3/2}\left(i-1,i,i+1\right)$ projects the total spin
of the three spins located at sites $i-1,i,i+1$ onto the 
$S=3/2$ subspace.
For an even number of sites $N$ and 
periodic BC, the ground state of the MG model (\ref{MG})
is two-fold degenerate and corresponds to
the two singlet states:
\begin{eqnarray}
|\Phi_1 \rangle &=& \left[1,2\right]  \left[3,4\right] ...  \left[N-1,N\right]
\nonumber \\
|\Phi_2 \rangle &=& \left[2,3\right]  \left[4,5\right] ...  \left[N,1\right] ,
\label{MGfond}
\end{eqnarray}
where $[i,i+1]$ denotes a singlet bond between
the spins at the $i$ and $i+1$ sites.
\setlength{\unitlength}{1.37cm}
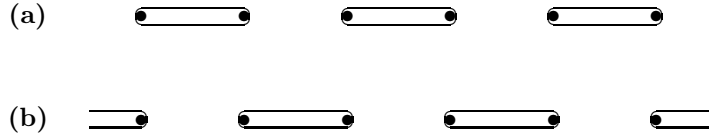
\begin{figure}[ht]
\begin{center}
\begin{picture}(5,3)
\multiput(0.1,1.4)(1,0){6}{\circle*{0.1}}
\multiput(0.1,2.4)(1,0){6}{\circle*{0.1}}
\put(0.6,2.4){\oval(1.1,0.16)}
\put(2.6,2.4){\oval(1.1,0.16)}
\put(4.6,2.4){\oval(1.1,0.16)}
\put(1.6,1.4){\oval(1.1,0.16)}
\put(3.6,1.4){\oval(1.1,0.16)}
\put(5.6,1.4){\oval(1.1,0.16)[l]}
\put(-0.4,1.4){\oval(1.1,0.16)[r]}
\put(-1.,2.4){\makebox(0,0){\bf (a)}} 
\put(-1.,1.4){\makebox(0,0){\bf (b)}}           
\end{picture}
\end{center}
\renewcommand{\baselinestretch}{0.8}
\caption{The two dimerized ground states at 
the Majumdar-Ghosh point; (a) $|\Phi_1 \rangle$, 
(b) $|\Phi_2 \rangle$.}
\label{figMG}
\end{figure}                        
The two ground states (\ref{MGfond}) are represented 
in Fig.  \ref{figMG}.
The lattice translation symmetry ${\cal T}_{a_0}$
is broken in the two ground states (\ref{MGfond}) and it
exchanges them:
${\cal T}_{a_0} |\Phi_{1,2} \rangle = |\Phi_{2,1} \rangle $.
The existence of gap in the MG model (\ref{MG}) 
has been shown rigorously by 
Affleck et al. \cite{aklt} and
the spin-spin correlation function can be determined exactly
at the  MG point \cite{shastry81}:
\begin{equation}
\langle {\vec S}_i \cdot {\vec S}_j \rangle = 
\frac{3}{4} \delta_{i,j} - \frac{3}{8} \delta_{|i-j|,1}.
\label{correlMG}
\end{equation}
This correlation function is thus zero for 
distance larger than one lattice spacing.
The ground-state degeneracy and the 
spontaneous breaking of a discrete Z$_2$ symmetry suggest 
the existence of topological excitations 
which interpolate between the two ground states (\ref{MGfond}).
These kink excitations can be viewed as 
the insertion of a spin-1/2 in a sea of singlet
valence bond states as shown in Fig. \ref{kinksing}.
This spin-1/2 excitation (spinon) is nothing but
\setlength{\unitlength}{2.17cm} 
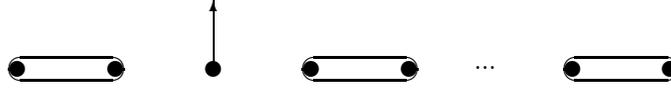
\begin{figure}
\begin{center}
\begin{picture}(4,1.0)
\multiput(0.4,0.4)(1.8,0){2}{\oval(0.7,0.14)}
\multiput(3.8,0.4)(1.8,0){1}{\oval(0.7,0.14)}
\multiput(0.1,0.4)(0.6,0){5}{\circle*{0.1}}
\multiput(3.5,0.4)(0.6,0){2}{\circle*{0.1}}
\put(1.3,0.4){\vector(0,1){0.42}}
\put(2.9,0.4){...} 
\end{picture}
\end{center}
\renewcommand{\baselinestretch}{0.8}
\caption{Dimerization kink (spinon) with $S=1/2$ quantum number.}
\label{kinksing}
\end{figure} 
a domain wall between the two states  (\ref{MGfond}).
For periodic BC, these kinks states appear always in pairs
(see Fig. \ref{2kinksing})
and the low-lying excitations of the MG model 
can be built starting from the state \cite{shastry81}:
\begin{eqnarray}
|p,m \rangle &=& \left[1,2\right] .. \left[2p -3, 2p -2 \right] 
\alpha_{2p -1} \left[2p, 2p +1 \right] ... \nonumber \\
&\;& \left[ 2m -2, 2m -1 \right] 
\alpha_{2m} \left[2m +1, 2m +2 \right] ... \left[N-1,N\right] ,  
\label{MGexc}
\end{eqnarray}   
where $\alpha_{2p -1}$ and $\alpha_{2m}$ denote spin-1/2 states
located at the sites $2p -1$ and $2m$ respectively.
A variational approach of the low-lying excitations \cite{shastry81,caspers84}
can then be done by considering a linear combination of 
states (\ref{MGexc}) and it leads to a triplet and singlet continuum 
spin excitations.
\setlength{\unitlength}{2.17cm} 
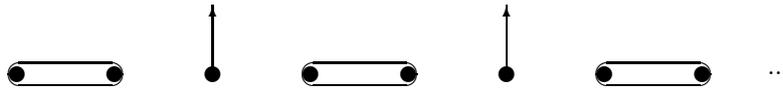
\begin{figure}
\begin{center}
\begin{picture}(5.,1.0)
\multiput(0.4,0.4)(1.8,0){2}{\oval(0.7,0.14)}
\multiput(4.0,0.4)(1.8,0){1}{\oval(0.7,0.14)}
\multiput(0.1,0.4)(0.6,0){6}{\circle*{0.1}}
\multiput(3.7,0.4)(0.6,0){2}{\circle*{0.1}}
\multiput(1.3,0.4)(1.8,0){2}{\vector(0,1){0.42}}
\put(4.7,0.4){...}
\end{picture}
\end{center}
\renewcommand{\baselinestretch}{0.8}
\caption{Spin-1 excitation built from two spinons.}
\label{2kinksing}
\end{figure}                              
In particular,
the dispersion relation of a massive spinon, obtained
within this approach, takes the form \cite{shastry81}:
\begin{equation}
\epsilon \left(k\right) = \frac{5J_1}{8}
+ \frac{J_1}{2} \; \cos \left( 2k a_0 \right) .
\label{shasexdis}
\end{equation}
This dispersion relation 
has been verified numerically by exact diagonalizations
\cite{sorensen98}.
Finally,
as shown by Caspers and Magnus \cite{caspers82}, 
there are additional exact singlet and triplet bound-states
at momentum $k=\pi/2 a_0$ which are degenerate with energy $E =J_1$.
Exact diagonalization and DMRG calculations \cite{sorensen98} 
have confirmed the existence of these bound-states 
for a small range of momenta close to $k=\pi/2a_0$
in the region $J_2 \ge J_1/2$.

In summary, the spontaneously dimerized phase for $J_2 > J_{2c}$ 
represents a distinct spin liquid phase stablized 
by frustration.
The main distinction, from the 
spin liquid phases of the two-leg spin ladder
and spin-orbital model, originates in the fractionalized
nature of the quantum number
carried by elementary excitations.
In particular,
instead of a sharp $S=1$ magnon peak as in the
two-leg spin ladder, 
the dynamical structure factor of the model in this dimerized phase
displays an incoherent background with additional bound-states
features.
The main difference between the dimerized
phase of  Fig. \ref{figMG} and the staggered dimerization 
phase of the spin-orbital model (\ref{hamso})
stems from the nature of the low-lying excitations.
In the latter phase, the excitations are 
a pair of propagating massive triplet kinks,
as described above,
whereas here the elementary excitations are massive spinons.

\begin{figure}[ht]
\begin{center}
\noindent
\epsfxsize=0.5\textwidth
\epsfbox{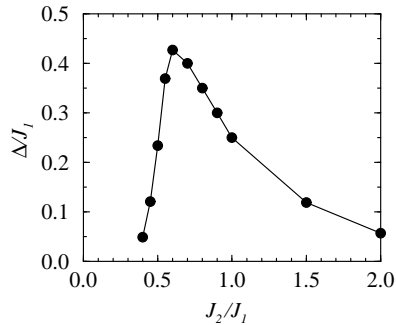}
\end{center}
\caption{\label{figspingap}%
Evolution of the spin gap, computed by DMRG, as a function 
of $J_2$ for $J_2 > J_{2c}$;
taken from Ref. 178. 
}
\end{figure}

This dimerized phase, with deconfined massive 
spinons excitations,
extends in  the entire region with
$J_2 > J_{2c}$ as it has been shown numerically.
Fig. \ref{figspingap} 
represents the evolution of the spin gap
$\Delta$, computed by DMRG \cite{whitebis96}, as function of the 
next-nearest neighbor interaction 
$J_2$.
As depicted by Fig. \ref{figspingap}, the spin gap is 
maximum for $J_2/J_1 \simeq 0.6$ and decreases to 
zero in the large $J_2$ limit.
The absence of a spin gap in this regime
can be easily understood since 
for $J_2 \gg J_1$, the $J_1-J_2$ spin chain can be better
viewed as a two-leg zigzag ladder (see Fig. \ref{zigagfig}).
In the limit $J_2 \rightarrow + \infty$, the two spin-1/2
Heisenberg chains are decoupled so that 
the model is critical: $\Delta =0$.
From the results of Fig. \ref{zigagfig},
it is very tempting to conclude that the whole region
with $J_2 > J_{2c}$ describes a single phase.
However, a subtle qualitative change occurs in the model
at the level of spin correlation functions
after the MG point in close parallel to the 
Haldane phase of the spin-1 bilinear-biquadratic chain (\ref{hambiqua})
just after the AKLT point (\ref{aklt}) \cite{joli96,fath00,nomura03}.
Indeed, an incommensurate behavior develops in 
the real-space spin-spin correlation function
for $J_2 > J_1/2$ 
\cite{chitra95,bursill95,whitebis96,watanabe99,aligia00}.
This incommensurability can be interpreted as 
a quantum signature of the spiral structure 
of the classical ground state of the $J_1-J_2$ spin chain
for $J_2 > J_1/4$.
After the MG point, 
the leading asymptotic of the spin-spin correlation function behaves as
\begin{equation}
\langle {\vec S}_i \cdot {\vec S}_{i+r} \rangle \sim 
\frac{1}{r^{1/2}} \cos \left(q \; r a_0\right)
\exp\left(-r a_0/\xi\right),
\label{corrinc}
\end{equation}
where the oscillation factor depends on $J_2/J_1$.
In parallel to the classification of 
commensurate-incommensurate transitions 
in Ising spin systems \cite{stephenson70},
the MG point, as the AKLT point (\ref{aklt}), 
is a disorder point of the first kind \cite{joli96}.
The momentum $q_{\rm max}$ which maximizes the 
static structure factor (Fourier transform 
of the spin-spin correlation function)
becomes incommensurate not at the MG point but slightly
after $J_{2 L} \simeq 0.5206 J_1 $ \cite{bursill95} 
at a Lifshitz point \cite{joli96}.
For $J_2 > J_{2 L}$, the static structure factor displays 
a double peak structure rather than a single peak.
The low-lying excitations are then characterized 
by an incommensurate momenta $q_{\rm max} \ne \pi/a_0$.
In contrast to the classical case,
it is worth noting that
the BKT phase transition between
the critical and dimerized phases  
and the onset of incommensurability at the Lifshitz point
occur at different points of the phase diagram,
$J_{2c} \simeq 0.2411 J_1$ and
$J_{2L} \simeq 0.5206 J_1$ respectively. 
Though no real phase transition occurs for $J_2 > J_{2c}$ 
in the model,  the dimerized phase in the region
$J_2 > J_{2L}$ can be distinguished 
from that at $J_{2c} < J_2 < J_{2L}$ due to the existence 
of this incommensurability.
In this respect, the spontaneously dimerized phase 
for $J_2 > J_{2L}$
with deconfined massive spinons
and incommensurate correlations represents a remarkable 1D 
spin liquid phase stabilized by frustration.

This incommensurability induced by
frustration is, in fact, quite general in 1D and not restricted 
to the spin-1/2 case. 
In particular, this phenomenon appears also in 
the spin-1 case.
The zero-temperature phase diagram of
the spin-1 $J_1-J_2$ Heisenberg
chain (\ref{j1j2ham}) has been investigated numerically 
\cite{tonegawa92,allen95,pati96,kolezhuk96,roth98,kolezhuk02}
and a spectral gap exists for all finite value of $J_2$.
A first-order transition occurs at 
$(J_2/J_1)_T \simeq 0.744$ \cite{kolezhuk96} separating
a Haldane phase from a double Haldane 
phase \cite{kolezhuk96,roth98,kolezhuk02}.
The quantum number carried by elementary excitations
is $S=1$ in full agreement with the semiclassical approach \cite{haldane00}
described above.
Frustration plays its trick here by giving rise to
a similar onset of incommensurability as in the spin-1/2 case.
A disorder point of the first kind occurs at
$(J_2/J_1)_D \simeq 0.284$ together with a  
Lifshitz point $(J_2/J_1)_L \simeq 0.3725$ \cite{kolezhuk96}
after which the static structure factor develops a 
two-peak structure.

\subsection{Field theory of spin liquid with incommensurate correlations}

Frustration represents thus a 
novel mechanism in one dimension 
for generating incommensurability as external 
magnetic fields or Dzyaloshinskii-Moriya interaction \cite{dm}.
The zigzag ladder 
representation (Fig. \ref{zigagfig}) of the 
$J_1-J_2$ model enables one
to investigate, by a weak coupling
approach $J_1 \ll J_2$,
the main characteristics of 
frustration and to shed light on the mechanism that 
gives rise to incommensurate correlations
in the large $J_2$ regime. 
In the spin-1/2 case, 
the onset of incommensurability and the presence 
of deconfined massive spinons might be understood 
starting from the limit where the two spin-1/2 AF Heisenberg
chains are decoupled.

The continuum limit of the $S=1/2$ two-leg zigzag ladder 
has been analysed by several groups
\cite{whitebis96,allen97,allen98,nersesyan98,cabra00,allen00,itoi01}.
The interacting part of the Hamiltonian density 
of the low-energy field theory reads as follows:
\begin{equation}
{\cal H}_{\rm int} \simeq
\; 
g_1 \left({\vec J}_{1 L} \cdot {\vec J}_{2 R}
+ {\vec J}_{2 L} \cdot {\vec J}_{1 R} \right)
+ g_2 \; {\vec n}_1 \cdot \partial_x {\vec n}_2 ,
\label{contzigzagsu2}
\end{equation}
where ${\vec J}_{a, L,R}$ ($a=1,2$) are the left-right su(2)$_1$
currents corresponding to the continuum description
of the ath spin-1/2 AF Heisenberg chain.
In Eq. (\ref{contzigzagsu2}),
${\vec n}_a$ denotes the staggered magnetization of 
the spin density of the chain with index $a=1,2$.
In the continuum limit, this ${\vec n}_a$ field  identifies
with the vector part of the primary field, with 
scaling dimension $\Delta_n =1/2$, of the 
su(2)$_1$ WZNW CFT transforming according to 
the fundamental representation of SU(2)
\cite{affleck85,affleck87,affleckhouches,bookboso}.
In this continuum description, frustration 
suppresses geometrically
the standard backscattering contribution
${\vec n}_1 \cdot {\vec n}_2$ which governs 
the low-energy physics of the two-leg spin ladder.
This term represents a strongly relevant perturbation
of scaling dimension $1$. 
It corresponds to
the energy operators of the underlying Ising models
of the low-energy description (\ref{2-legham}) of the two-leg spin
ladder \cite{schulz86,shelton96}.
As seen in Section 2, this backscattering contribution 
leads to the formation of a spin gap in this
model as well as the confinement of the original massless
spinons into optical magnons.
In presence of frustration, 
the interacting part (\ref{contzigzagsu2}) is now 
only marginal relevant and consists of two terms
of different nature.
The first one, with coupling constant $g_1$, 
is a current-current interaction similar 
to that which appears in the effective field 
theory (\ref{curcurham}) of the spin-1/2
$J_1-J_2$ Heisenberg chain in the weak coupling 
limit $J_2 \ll J_1$.
The second marginal contribution in Eq. (\ref{contzigzagsu2}),
called twist term \cite{nersesyan98},
${\cal O}_{\rm twist} = {\vec n}_1 \cdot \partial_x {\vec n}_2$
is a novel parity-breaking perturbation which
contains the staggered magnetizations of each chain
and also a spatial derivative.
In contrast to the current-current perturbation, 
this twist term is not a scalar under the Lorentz 
transformation in $1+1$ dimensions but behaves
as a vector. In the CFT jargon, this kind of 
perturbation is characterized by a non-zero conformal spin $S=\pm1$
together with its scaling dimension $\Delta=2$.
In fact, this twist perturbation has been forgotten 
in the first bosonization analysis of the two-leg zigzag
spin ladder \cite{whitebis96,allen97}.
It has been discovered by Nersesyan et al. \cite{nersesyan98} 
and independently by Allen \cite{allen98}.
It is interesting to observe that a similar operator appears in the 
effective field theory approach to the spin-1 two-leg
zigzag ladder in the large $J_2$ limit \cite{allenbis00}.
This twist perturbation is thus the hallmark of frustration 
in the low-energy description of frustrated spin ladders
with zigzag interchain interaction.

The effect of such a non-zero conformal spin perturbation 
is non-trivial since the usual irrelevant versus relevant 
criterion of perturbative field theory does not hold 
for such a non-scalar contribution (see for instance
the discussion in the book \cite{bookboso}).
In the case of a $S = \pm 1$ perturbation,
it is expected that the generic effect of this contribution is the onset
of incommensurability.
In this respect, a simple example of incommensurability 
arising from such a perturbation is
the spin-1/2 XXZ Heisenberg chain in a magnetic field 
along the z-axis. The Hamiltonian density of this model,
obtained within the bosonization approach, reads as 
follows \cite{chitra97}:
\begin{equation}
{\cal H} =
\frac{v}{2} \left[ \left(\partial_x \Phi \right)^2
 + \left(\partial_x \Theta \right)^2  \right]
  - g \cos \beta \Phi  -  h \;
\partial_x \Phi .
\label{sinegordonchamp}
\end{equation}
Here the $S=\pm 1$ perturbation is described by the 
uniform part of the spin density 
$\partial_x \Phi$.
It is well known that in this model, when 
the magnetic field $h$ is increased, 
a commensurate-incommensurate phase transition takes place
with the appearance of an incommensurate phase with critical correlation 
functions \cite{nersesyan78,pokrovsky79,schulz80,okwamoto80}.
This transition occurs for a finite or vanishing 
magnetic field depending on the relevance or irrelevance
of the $\cos \beta \Phi$ term.
A similar incommensurability generated by a $S=\pm 1$
conformal spin perturbation occurs in the
spin-1/2 XXZ Heisenberg chain in a transverse magnetic field 
\cite{nersesyan93,dmitriev02,dutta03}, 
in a model of two spinless Luttinger chains 
weakly-coupled by a single-particle interchain hopping
\cite{nersesyan91,nersesyan93,fendley01}, 
and in the quantum axial next-to-nearest neighbor Ising (ANNNI)
chain in a transverse magnetic field \cite{annni01}.
Finally, it is important to note that 
there exists some exactly solvable CFT models perturbed by 
$S=\pm 1$ conformal spin term for which 
the presence of incommensurability can be shown
non-perturbatively \cite{cardy93,tsvelik01}.

All these results suggest that the twist term
${\cal O}_{\rm twist} = {\vec n}_1 \cdot \partial_x {\vec n}_2$,
as proposed by Nersesyan et al. \cite{nersesyan98},
should be at the origin 
of the incommensurability found numerically in the
spin-1/2 $J_1-J_2$ Heisenberg chain (\ref{j1j2ham}) 
when $J_2 > 0.52 J_1$.
To this end,
this twist term and the current-current interaction
of Eq. (\ref{contzigzagsu2})
can be expressed in terms of the four 
Majorana fermions $\xi_{R,L}^a, a=0,1,2,3$
of the continuum limit (\ref{2-legham}) of the standard
two-leg spin ladder \cite{allen98,nersesyan98}.
In particular,
the current-current interaction is built from
of the following tensor:
${\cal O}^{a b}_{\rm cc} = \xi_R^a \xi_L^a \xi_R^b \xi_L^b$
with $a,b = 0,1,2,3$ and $a \ne b$.
The twist perturbation is also local in terms 
of these Majorana fermions but with a different structure.
A typical term that enters its expression is
${\cal O}^{a b c d}_{\rm twist} = \xi_R^a \xi_L^b \xi_L^c \xi_L^d$
with $a,b,c,d =0,1,2,3$ and $a\ne b \ne c \ne d$.
The non-zero conformal spin of this twist term
is reflected here by the different number of 
right and left fermions in this expression.
The field theory (\ref{contzigzagsu2}), expressed in terms
of these four Majorana fermions, turns out not to be integrable.
The renormalization group (RG) flow analysis reveals that the
current-current interaction and the twist term are equally
important in the IR limit. They reach a strong coupling
regime simultaneously with a fixed ratio \cite{allen98,allen00,itoi01}.
The nature of this strong coupling regime is still 
an open problem since the emerging IR field theory 
is not integrable and one cannot disentangle the effect of
the two interactions of Eq. (\ref{contzigzagsu2}).
However, it is very tempting to explain
the main characteristics of the spontaneously dimerized phase 
with incommensurate correlation of the 
spin-1/2 $J_1-J_2$ Heisenberg chain in the large $J_2$ limit from 
these two contributions. On one hand,
the current-current contribution of Eq. (\ref{contzigzagsu2}),
equivalent to two decoupled SU(2) Thirring model,
has similar properties as the field theory
(\ref{curcurham}) obtained in the weak 
coupling regime $J_2 \ll J_1$.
The spontaneously dimerization and the existence of massive deconfined 
spinons should result from this interaction.
In fact,
this current-current interaction appears alone,
without a twist term, in the continuum limit
of a frustrated two-leg spin ladder with crossings
along a special line of the couplings \cite{allen00}.
It has been shown that this model, in the weak coupling limit,
is characterized by a weak spontaneously dimerization 
with massive spinons as elementary excitations \cite{allen00}.
Finally, as already stressed, 
the twist perturbation of Eq. (\ref{contzigzagsu2})
should be at the origin of the incommensurability 
in the large $J_2$ limit \cite{nersesyan98}.
In this respect,
the effect of the twist term of the two-leg
zigzag ladder has been analysed 
within a RPA approach
and it leads indeed to some incommesurability behavior in the 
spin-spin correlation function \cite{allenpriv}.

A simple way to disentangle the effects 
of the current-current interaction and the twist
perturbation is to introduce an exchange anisotropy
which makes the twist contribution more relevant 
in the RG sense \cite{nersesyan98}.
The extreme case is the XY version of the spin-1/2
$J_1-J_2$ Heisenberg chain with Hamiltonian:
\begin{equation}
{\cal H}_{\rm XY} = J_1 \sum_n \left( S_n^x S_{n+1}^x + 
S_n^y S_{n+1}^y\right) + 
J_2 \sum_n \left( S_n^x S_{n+2}^x + 
S_n^y S_{n+2}^y\right).
\label{xyhamj1j2}
\end{equation}
In the ladder limit when $J_2 \gg J_1$, 
the low-energy physics of this model can be 
analysed by means of the bosonization approach.
Introducing two bosonic fields $\Phi_{\pm}$,
the resulting bosonic Hamiltonian density of the model
reads as follows \cite{nersesyan98}:
\begin{equation}
{\cal H}_{\rm XY} \simeq \frac{v}{2}\sum_{a=\pm}\left(\left(
\partial_x \Phi_a\right)^2
+ \left(\partial_x \Theta_a\right)^2\right)
+ g \; \partial_x
\Theta_+ \sin\left(\sqrt{2 \pi}{\Theta}_-\right),
\label{contzigzagxy}
\end{equation}
where $\Theta_{\pm}$ are the dual fields associated
to the bosonic fields $\Phi_{\pm}$.
The Hamiltonian (\ref{contzigzagxy}) describes a nontrivial
field theory with a relevant 
$S=\pm 1$ conformal spin twist perturbation
with scaling dimension $\Delta_g = 3/2$. 
In this strong anisotropic XY case,
the current-current perturbation 
is less dominant and can be safely neglected 
to derive the nature of the phase of the model 
when $J_2 \gg J_1$.
The presence of incommensurability in the system
can then be found using a mean-field analysis of the model
by decoupling the two pieces of the 
twist term of Eq. (\ref{contzigzagxy}). 
In particular, the leading asymptotic 
behavior of the transverse spin-spin correlation 
functions of the model obtained by this mean-field 
approach is given by \cite{nersesyan98}:
\begin{equation}
\langle S_1^{\dagger}\left(x\right) S^{-}_a\left(0\right) \rangle \sim
\frac{\exp\left(iq x\right)}{|x|^{1/4}}, \; \; a=1,2 ,
\label{correlzigzagS}
\end{equation}
which displays an incommensurate critical behavior with
an oscillating factor: 
$q - \pi/a_0 \sim (J_1/J_2)^{2}$.
The physical picture that emerges from this mean-field
analysis is the existence of a critical spin nematic 
phase that preserves the U(1) and time-reversal 
symmetries and displays long-range 
chiral ordering in its ground state:
$\langle ({\vec S}_{n+1} \wedge {\vec S}_{n})_z \rangle \ne 0$.
It is important to note that this spin nematic phase
does not break the time-reversal symmetry but spontaneously breaks
a $Z_2$ symmetry of the model which
is a tensor product of a site-parity and link-parity
symmetries on the two chains: $P_L^{(1)} \times P^{(2)}_S$. 
The z-component of the spin current $J^{z}_{as}$ associated
to the ath spin-1/2 XY chain ($a=1,2$) takes a non-zero
expectation value in the ground state of the 
mean field Hamiltonian \cite{nersesyan98}: 
\begin{equation}
\langle J^{z}_{1s} \rangle = 
\langle J^{z}_{2s} \rangle = 
- \frac{v}{\sqrt{ 2 \pi}} \;
\langle \partial_x \Theta_{+} \rangle \ne 0.
\label{spincurr}
\end{equation}
As a result, 
this produces a picture of local nonzero
spin currents polarized along the z-anistropy axis circulating 
around the triangular plaquettes
of the two-leg zigzag spin ladder.     
This phase has been found numerically
by Nishiyama \cite{nishiyama00},
who has investigated the existence of chiral
order of the Josephson ladder with half a flux quantum per plaquette, 
and also by Hikihara et al. \cite{hikihara01}
by means of the DMRG approach.
This critical incommensurate phase 
is the quantum analogue of the classical spiral 
phase with chiral ordering \cite{villain77}. 
It is thus natural to expect that this
phase is not restricted to the spin-1/2
case but should exist in the range of parameters
of the model (\ref{xyhamj1j2}) in the general spin case.
Indeed, the existence of this critical spin nematic phase
in the spin-$S$ case has been shown 
numerically for $S=1,3/2,2$ \cite{hikihara01,kaburagi99},
by means of a semiclassical method \cite{kolezhuk00},
and finally by the Abelian bosonization approach \cite{phlejoli01}
of a general spin $S$ introduced by Schulz \cite{schulz86}.

\subsection{Extended criticality stabilized by frustration}

In addition to incommensurability,
a second characteristic of classical canted magnets
is the presence of three gapless spin-wave modes
instead of two as in  the N{\'e}el colinear state.
An interesting question is whether frustration
can lead to new type of 
emerging quantum criticality not encountered 
in unfrustrated spin chains or ladders.
In this last case, the critical behavior of
half-integer AF Heisenberg spin chains
or odd-legged spin ladders
is characterized by one gapless bosonic mode
i. e. by a CFT with central charge $c=1$.
As it has already been pointed out in the previous
section, a striking effect of frustration 
in the continuum limit stems from the fact that the low-energy 
effective field theory is mainly governed by marginal interactions.
New type of IR critical behaviors may result from the delicate
balance between these marginal perturbations.
In the following, we shall give some examples 
of critical phases and quantum critical points
with extended criticality stabilized by frustration 
in 1D spin systems.

\subsubsection{Critical phases with SU(N) quantum criticality}

A first example of a system with extended criticality
is the 1D spin-orbital model (\ref{hamso}) in 
a regime where frustration reveals itself $K \sim J$.
At the special point $J= K/4$, 
the Hamiltonian (\ref{hamso}) can be expressed in terms
of a product of two-body permutation operator
in $S_1$ and $S_2$ subspaces:
\begin{eqnarray}
{\cal H}_{\rm so}  &=&
J \sum_i \left(2 {\vec S}_{1,i} \cdot {\vec S}_{1,i+1} + \frac{1}{2}
\right)
\left(2 {\vec S}_{2,i} \cdot {\vec S}_{2,i+1} + \frac{1}{2}
\right) \nonumber \\
&=& J \sum_i P_{i,i+1}^{(S_1=1/2)} \; P_{i,i+1}^{(S_2=1/2)}.
\label{hamsosu4}
\end{eqnarray}
Since this Hamiltonian exchanges both ${\vec S}_{1,i}$ and 
${\vec S}_{2,i}$ spins at the same time, the spin-orbital model 
at $J= K/4$ is not only SU(2) $\times$ SU(2)
symmetric but actually has an elarged SU(4) symmetry  which unifies
the spin and orbital 
degrees of freedom \cite{li98,yamashita98,mila99}.
More precisely, the Hamiltonian (\ref{hamsosu4})
can be recasted as an AF Heisenberg spin chain 
with SU(4) spins up to a constant:
\begin{equation}
{\cal H}_{\rm SU(4)} = J \sum_i \sum_{A=1}^{15}
T_i^A T_{i+1}^A ,
\label{hamsu4}
\end{equation}
where $T^A$  are the 15 generators
belonging to the fundamental representation of SU(4).
This model is exactly solvable by means of 
the Bethe ansatz \cite{sutherland75}
and its low-energy spectrum consists
of three gapless spinons with wave vectors
$\pm \pi/2a_0$ and $\pi/a_0$ \cite{sutherland75,li99}.
As shown by Affleck \cite{affleck88},
the critical theory corresponds 
to the su(4)$_1$ WZNW model with central
charge $c=3$ (three massless bosonic modes).
The existence of this quantum critical point 
with a SU(4) symmetry allows us to study 
the spin-orbital model (\ref{hamso})
by a continuum description in an intermediate coupling regime $J \sim K$
where frustration shows off.
In this respect, it is important to notice
that, according to the Zamolodchikov's c theorem \cite{zamolo86},
the SU(4) critical point with central charge $c=3$
cannot be reached by a RG trajectory 
starting from the decoupling limit ($K=0$) of two spin-1/2
AF Heisenberg chains with total central charge $c=2$.
Stated differently, the physics in the neighborhood 
of the SU(4) point cannot be understood in terms
of weakly coupled $S=1/2$ Heisenberg chains.
The strategy to tackle with this intermediate coupling
regime is to start from the 
SU(4) Hubbard chain at quater-filling
and apply the bosonization approach to
obtain the continuum description of 
the spin densities ${\vec S}_{1,2}(x)$ at the SU(4) point
in the large Coulomb repulsion limit.
The low-energy field theory which describes 
small deviations from the SU(4) symmetric point
can then be derived \cite{phleso99,itoi00}.
The resulting effective Hamiltonian 
density associated to the symmetry breaking scheme
SU(4) $\rightarrow$ SU(2) $\times$ SU(2)
of the spin-orbital model (\ref{hamso}) 
reads as follows \cite{phleso99}:
\begin{eqnarray}
{\cal H}_{\rm eff} &=& -\frac{i v_s}{2}
\left({\vec \xi}_{sR} \cdot \partial_x {\vec \xi}_{sR}
- {\vec \xi}_{sL} \cdot \partial_x {\vec \xi}_{sL} \right)
 -\frac{i v_o}{2}
 \left({\vec \xi}_{oR} \cdot \partial_x {\vec \xi}_{oR}
 - {\vec \xi}_{oL} \cdot \partial_x {\vec \xi}_{oL} \right)
 \nonumber \\
&+& g_1 \left({\vec \xi}_{sR} \cdot {\vec \xi}_{sL}\right)^2
+ g_2 \left({\vec \xi}_{oR} \cdot {\vec \xi}_{oL}\right)^2
+ g_3 \left({\vec \xi}_{sR} \cdot {\vec \xi}_{sL}\right)
\left({\vec \xi}_{oR} \cdot {\vec \xi}_{oL}\right),
\label{hamsocont}
\end{eqnarray}
where ${\vec \xi}_{sR,L}$ and ${\vec \xi}_{oR,L}$
are two triplet of Majorana fermions which
act respectively in the spin and orbital sectors.
In Eq. (\ref{hamsocont}), we have considered 
a more general situation than the model (\ref{hamso})
by allowing the exchange in spin ($J_1$) and 
orbital ($J_2$) channels to be different:
the three coupling constants $g_i$ are independent.
The interaction of the low-energy field theory 
(\ref{hamsocont}) is marginal and describes two
SO(3) Gross-Neveu \cite{grossneveu74} models
marginally coupled.
This field theory is not integrable so that 
one has to recourse to perturbation theory to 
elucidate the phase diagram of the spin-orbital
model in the vicinity of the SU(4) symmetric point.
In fact,  the all-order beta functions 
of the field theory (\ref{hamsocont}) can be 
determined using the approach of Gerganov et al. \cite{leclair01}.
These authors have computed the all-order beta functions of 
a general model with anisotropic current-current interactions
in a special minimal scheme prescription.
The model (\ref{hamsocont}) can be viewed 
as a current-current interaction 
corresponding to the symmetry breaking scheme 
SO(6) $\rightarrow$ SO(3)$\times$ SO(3).
The application of the general formula given in Ref. \cite{leclair01} 
confirms the conclusions derived from the one-loop
calculation \cite{phleso99,itoi00}
which reveals the existence of two phases 
with different remarkable properties.

A first phase, for $J_1 \simeq J_2 > K/4$,
has a spectral gap and the ground state 
has a similar staggered dimerization as 
the weak-coupling phase of Fig. \ref{dimerpi}.
Provided the anisotropy is not too large $J_1 \ne J_2$
in the vicinity of the SU(4) symmetric point,
the massive phase displays an approximate SO(6) $\sim$ SU(4)
enlarged symmetry.
Indeed, by neglecting the velocity anisotropy
in Eq. (\ref{hamsocont}), 
the RG equations reveal a flow 
to strong coupling with an 
attraction along a special direction
with enlarged symmetry described by the following
Hamiltonian density \cite{phleso99}:
\begin{equation}
{\cal H}_{\rm IR} \simeq - \frac{iv_{*}}{2}
\sum_{a=1}^{6}\left(\xi_{R}^a \partial_x \xi_{R}^a
- \xi_{L}^a \partial_x \xi_{L}^a\right)
+ g_{*} \left(\sum_{a=1}^{6} \xi_{R}^a \xi_{L}^a \right)^2,
\label{so6GN}
\end{equation}
with $g_{*} >0$.
This Hamiltonian is identified as the SO(6) Gross-Neveu 
model which is a massive integrable field theory.
Its spectrum is known and consists of the fundamental
fermion, with mass $M$, together with a 
kink and anti-kink with mass 
$m_{\rm kink} = M/\sqrt{2}$ \cite{zamolo79}.
The initial model (\ref{hamsocont}), SU(2) $\times$ SU(2) symmetric,
acquieres thus in the IR limit
an enlarged SO(6) symmetry. 
In more physical terms,
the spin and orbital degrees of freedom are unified
and described by a same multiplet with six components.
A similar example of symmetry restauration by
interactions is the emergence of a SO(8) symmetry 
in weakly-coupled two-leg Hubbard ladder at half-filling
\cite{lin98} and in the SU(4) Hubbard chain at half-filling
\cite{hubsu4}.

The second phase with $J_1 \simeq J_2 < K/4$
has striking properties.
All couplings in Eq. (\ref{hamsocont}) flow to
zero in the IR limit and the interaction is marginal irrelevant.
The six Majorana fermions are thus massless and
the phase displays extended quantum criticality
characterized by a central charge $c=3$.
Spin-spin correlation functions decay algebraically
with exponent $3/2$ and exhibit a four-site periodicity
($2k_F =\pi/2 a_0$).         
\begin{figure}[ht]
\begin{center}
\noindent
\epsfxsize=0.5\textwidth
\epsfbox{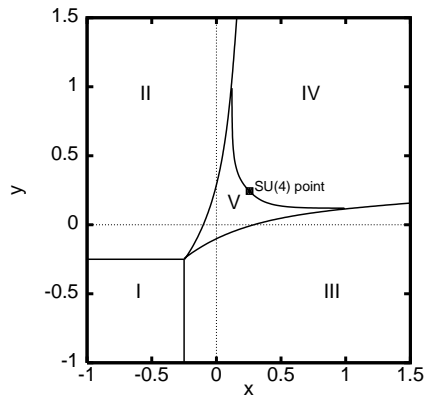}
\end{center}
\caption{\label{sodiagphas}%
Phase diagram of the spin-orbital model obtained  
by DMRG; $x = J_2/K$ and $ y = J_1/K$;
taken from Ref. 231.
}
\end{figure}
The critical behavior at the SU(4) point (\ref{hamsu4})
extends to a finite region of the 
phase diagram of the spin-orbital model.
In this respect, this remarkable gapless phase with extended
quantum criticality $c=3$
represents a new universality class, stabilized 
by frustration, in spin chains and spin ladders.
It is worth noting that this phase has 
a maximum of gapless modes allowed by 
the classical structure of a spiral (three Goldstone
modes). 

The phase diagram of the anisotropic spin-orbital model (\ref{hamso})
with  $J_1 \ne J_2$ has been investigated numerically
by means of the DMRG technique \cite{pati98,yamashitabis00,itoi00}.
Figure \ref{sodiagphas} represents the resulting
$T=0$ phase diagram obtained in Refs. \cite{yamashitabis00,itoi00}.
In phase I, both spin and orbital degrees of freedom 
are in fully polarized ferromagnetic states.
In phase II, the orbital degres of freedom are still  
in a ferromagnetic state whereas the spin degrees of freedom
are now critical
(criticality of a spin-1/2 AF Heisenberg chain)
and vice versa in phase III. Phase IV corresponds to 
the staggered dimerized phase of Fig. \ref{dimerpi}
while phase V is the $c=3$ gapless phase with extended quantum
criticality. 
In fact, this critical phase has been first discovered
only along a special line $-K/4 < J_1 = J_2 < K/4$  
in the DMRG calculation of Pati et al. \cite{pati98}.
Additional DMRG works \cite{yamashitabis00,itoi00} 
have shown the existence of this gapless phase 
in an extended region (Phase V) of 
the phase diagram (Fig. \ref{sodiagphas}).

A second model which displays extended quantum criticality 
induced by frustration is the spin-1 bilinear-biquadratic
chain (\ref{hambiqua}) with
a biquadratic coupling constant $\beta >1$.
In this regime, frustration manifests itself and 
the classical ground state is an incommensurate spiral.
The Haldane phase of the spin-1 bilinear-biquadratic chain (\ref{hambiqua})
with $|\beta| < 1$ ends at the 
integrable point with $\beta =1$, the so-called 
Uimin-Lai-Sutherland \cite{uimin70,sutherland75} point.
This model can be expressed in terms of SU(3) spins
and takes the form of 
a SU(3) AF Heisenberg spin chain:
\begin{equation}
{\cal H}_{\rm SU(3)} = J \sum_i \sum_{A=1}^{8}
T_i^A T_{i+1}^A ,
\label{hamsu3}
\end{equation}
where $T^A$  are the 8 generators
belonging to the fundamental representation of SU(3).
This model is exactly solvable by means of
the Bethe ansatz \cite{uimin70,sutherland75}
and its low-energy spectrum consists
of two gapless spinons with wave vectors
$\pm 2\pi/3 a_0$ \cite{sutherland75}.
As shown by Affleck \cite{affleck88},
the critical theory corresponds
to the su(3)$_1$ WZNW model with central
charge $c=2$ (two massless bosonic modes).
The existence of this quantum critical point
with a SU(3) symmetry allows to study
the spin-1 bilinear-biquadratic chain (\ref{hambiqua})  
in the vicinity of $\beta \simeq 1$ by means of a
field theory approach \cite{itoi97}.
The low-energy effective field theory is described
by marginal current-current interactions 
associated to the symmetry breaking scheme 
SU(3) $\rightarrow$ SU(2) \cite{itoi97}.
The one-loop RG flow near $\beta \simeq 1$ 
reveals the existence of two distinct phases.
On one hand, a phase with
a dynamical mass generation when $\beta < 1$
which signals the onset of the Haldane phase.
On the other hand, a massless phase for $\beta > 1$
where interactions are marginal irrelevant.
The SU(3) quantum criticality of the Uimin-Lai-Sutherland point,
with two gapless bosonic modes,
extends to a finite region of the phase diagram
of the generalized spin-1 Heisenberg chain (\ref{hambiqua}).
This massless phase has been first predicted in a
numerical investigation of this model 
by F{\'a}th and S{\'o}lyom \cite{fath91}.
This massless phase with approximate SU(3) symmetry, stabilized 
by frustration, is 
similar in spirit to the previous SU(4) massless phase of the 
spin-orbital model (\ref{hamso}).
Finally, it is worth noting that this $c=2$ massless
phase appears also in the phase diagram
of the so-called spin tube model, a three-leg spin ladder
with frustrated periodic interchain interaction,
in a magnetic field \cite{citro00}.

\subsubsection{Chirally stabilized critical spin liquid}

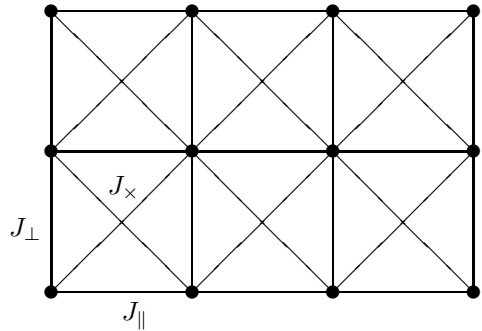
\begin{figure}[h]
\begin{center}
\setlength{\unitlength}{1.87cm}
\begin{picture}(4,1)
\multiput(0.5,0.4)(1,0){4}{\circle*{0.1}}
\multiput(0.5,1.4)(1,0){4}{\circle*{0.1}}
\multiput(0.5,2.4)(1,0){4}{\circle*{0.1}}
\drawline[10](0.5,0.4)(1.5,0.4)
\drawline[10](1.5,0.4)(2.5,0.4)
\drawline[10](2.5,0.4)(3.5,0.4)
\drawline[10](0.5,1.4)(1.5,1.4)
\drawline[10](1.5,1.4)(2.5,1.4)
\drawline[10](2.5,1.4)(3.5,1.4)
\drawline[10](0.5,2.4)(1.5,2.4)
\drawline[10](1.5,2.4)(2.5,2.4)
\drawline[10](2.5,2.4)(3.5,2.4)
\drawline[10](0.5,0.4)(0.5,1.4)
\drawline[10](1.5,0.4)(1.5,1.4)
\drawline[10](2.5,0.4)(2.5,1.4)
\drawline[10](3.5,0.4)(3.5,1.4)
\drawline[10](0.5,1.4)(0.5,2.4)
\drawline[10](1.5,1.4)(1.5,2.4)
\drawline[10](2.5,1.4)(2.5,2.4)
\drawline[10](3.5,1.4)(3.5,2.4)
\drawline[10](0.5,0.4)(1.5,1.4)
\drawline[10](0.5,1.4)(1.5,0.4)
\drawline[10](0.5,2.4)(1.5,1.4)
\drawline[10](0.5,1.4)(1.5,2.4)
\drawline[10](1.5,0.4)(2.5,1.4)
\drawline[10](1.5,1.4)(2.5,0.4)
\drawline[10](1.5,2.4)(2.5,1.4)
\drawline[10](1.5,1.4)(2.5,2.4)
\drawline[10](2.5,0.4)(3.5,1.4)
\drawline[10](2.5,1.4)(3.5,0.4)
\drawline[10](2.5,2.4)(3.5,1.4)
\drawline[10](2.5,1.4)(3.5,2.4)
\put(1,0.2){$J_{\parallel}$}
\put(0.2,0.8){$J_{\perp}$}
\put(0.9,1.1){$J_{\times}$}
\end{picture}
\end{center}
\renewcommand{\baselinestretch}{0.5}
\caption{Three-leg spin ladder with crossings.}
\label{fig3legfrus}
\end{figure}

Frustration can also induce an exotic quantum critical point 
in frustrated spin ladders.
A possible lattice realization of this phenomenon 
is a model (see Fig. \ref{fig3legfrus}) of three $S=1/2$ AF
Heisenberg spin chains weakly coupled by on-rung 
$J_{\perp}$ and plaquette-diagonal $J_{\perp}$ interchain 
interactions \cite{phlechir98}.
The Hamiltonian of this model reads as follows:
\begin{eqnarray}
{\cal H}_{\times} = &J_{\parallel}& \sum_i \sum_{a=1}^{3} {\vec S}_{a,i}
\cdot {\vec S}_{a,i+1} +
J_{\perp}  \sum_i {\vec S}_{2,i}\cdot \left({\vec S}_{1,i}
+ {\vec S}_{3,i} \right) \nonumber \\
&+& J_{\times} \sum_i
\left[
\left({\vec S}_{1,i} + {\vec S}_{3,i}\right)
\cdot {\vec S}_{2,i+1}
+ \left({\vec S}_{1,i+1} + {\vec S}_{3,i+1}\right)
\cdot {\vec S}_{2,i}\right].
\label{3legfrus}
\end{eqnarray}
\vspace{2.5cm}
In the continuum limit with $J_{\perp}, J_{\times} \ll J_{\parallel}$,
the Hamiltonian of the lattice model (\ref{3legfrus})
takes the form:
\begin{eqnarray}
{\cal H}_{\times} &=& \frac{2\pi v}{3} \sum_{a=1}^{3}
\left({\vec J}_{aR}^2 + {\vec J}_{aL}^2 \right)
+ {\tilde g} \; {\vec n}_{2}
\cdot \left({\vec n}_1 +  {\vec n}_3 \right)
\nonumber \\
&+&  g \; \left[{\vec J}_{2 R} \cdot \left({\vec J}_{1 L} +
{\vec J}_{3 L} \right) +
{\vec J}_{2 L} \cdot \left({\vec J}_{1 R} +
{\vec J}_{3 R} \right) \right],
\label{conintmod1}
\end{eqnarray}
with $g = a_0 (J_{\perp} + 2 J_{\times})$
and ${\tilde g} = a_0\left(J_{\perp} - 2 J_{\times}\right)$.
In Eq. (\ref{conintmod1}),
${\vec J}_{aR,L}$ ($a=1,2,3$) are
the right and left chiral su(2)$_1$ currents
which accounts for the low-energy description of
the uniform part of the spin density of the
ath spin-1/2 chain.                                               
It is interesting to note that in Eq. (\ref{conintmod1}), 
there is no 
marginally relevant twist perturbation ${\vec n}_1 
\partial_x {\vec n}_2$ which appears 
in the continuum description of the two-leg 
spin ladder with a small zigzag interchain 
coupling (\ref{contzigzagsu2}).
The two interaction terms in Eq. (\ref{conintmod1})
are of different nature.
On one hand, the first contribution, with coupling
constant ${\tilde g}$, is 
a strongly relevant perturbation with  
scaling dimension $\Delta_{{\tilde g}} = 1$. On the other hand, 
the second term is
a current-current interaction
which is only
marginal and, as long as
${\tilde g}$ is not too small, can be discarded.
As a result, for generic values of
$g$ and ${\tilde g}$, the low-energy physics of the 
model (\ref{3legfrus}) will be essentially that of the
standard three-leg ladder, and frustration will
play no role
(except for renormalization of mass gaps and
velocities).
The important point here
is that in contrast with non-frustrated ladders,  the two coupling
constants $g,{\tilde g}$ can vary independently, and
there exists a vicinity of the 
line $J_{\perp}= 2 J_{\times}$ (${\tilde g} = 0$)
where the low-energy properties of the model are mainly determined
by current-current interchain interaction.

Along the special line $J_{\perp}= 2 J_{\times}$,
the low-energy effective Hamiltonian (\ref{conintmod1})  simplifies 
as follows \cite{phlechir98}:
\begin{eqnarray}
{\cal H}_{\times} =
&-& \frac{iv}{2} \left(\xi_{R}^{0} \partial_x  \xi_{R}^{0} -
\xi_{L}^{0} \partial_x  \xi_{L}^{0}  \right)
+\frac{2\pi v}{3}
\left({\vec J}_{2R}^2 + {\vec J}_{2L}^2 \right)
\nonumber \\
&+& \frac{\pi v}{2}
\left({\vec I}_{R}^2 + {\vec I}_{L}^2 \right)
+  g \; \left[{\vec J}_{2R} \cdot {\vec I}_{L}
+ {\vec J}_{2L} \cdot {\vec I}_{R} \right],
\label{conintmod2}
\end{eqnarray}
where $\xi_{R,L}^{0}$ are Majorana fermions which 
are associated to the Z$_2$ ($1 \rightarrow 3$) 
discrete interchange symmetry between the surface chains
labelled $a=1,3$ in the lattice 
model (\ref{3legfrus});
the total chiral current of the surface chains is noted
${\vec I}_{R,L} = {\vec J}_{1R,L} + {\vec J}_{3R,L}$
and corresponds to a su(2)$_2$ WZNW current.
The Hamiltonian (\ref{conintmod2}) has an interesting structure.
First, the Majorana fermions $\xi_{R,L}^{0}$
do not participate in the interaction and 
remain thus critical in the IR limit.
These massless degrees of freedom can be interpreted as 
an effective two-dimensional Ising model at $T = T_c$
which accounts for singlet excitations between the surface
chains.
All non-trivial physics of the model is incorporated
in the current-current dependent part of the Hamiltonian
(\ref{conintmod2}), denoted by ${\cal H}_{cc}$, 
which describes marginally coupled
su(2)$_1$ and su(2)$_2$ WZNW models.
Moreover, this Hamiltonian separates into two commuting and
{\it chirally asymmetric} parts:
${\cal H}_{cc} = {\cal H}_1 + {\cal H}_2, ~([{\cal H}_1, {\cal H}_2] = 0)$,
where
\begin{equation}
{\cal H}_1 =  \frac{\pi v}{2} \;  {\vec I}_{R}^2  +
 \frac{2\pi v}{3} \; {\vec J}_{2L}^2
 + g \;  {\vec I} _{R} \cdot {\vec J}_{2L},
\label{hamsep}
\end{equation}
and ${\cal H}_2$ is  obtained from ${\cal H}_1$ by
inverting chiralities of all the currents.
The model (\ref{hamsep}) is integrable by means of 
the Bethe-ansatz approach \cite{polyakov83,andrei98}.
A simple RG analysis shows that, at $g > 0$,
the interaction is marginally
relevant. Usually the development of
a strong coupling regime is accompanied by a
dynamical mass generation and the loss of conformal
invariance in the strong coupling limit. 
However, here 
due to the chiral asymmetry of ${\cal H}_1$,
it turns out that the effective interaction
flows towards an intermediate fixed point
where conformal invariance is recovered with
a smaller central charge.
This critical behavior has been identified as 
the universality class of chirally stabilized
fluids, introduced by Andrei, Douglas,
and Jerez \cite{andrei98}.
In the case of the model (\ref{hamsep}), 
the symmetry of the IR fixed point,
obtained from the Bethe-ansatz analysis \cite{andrei98}, 
turns out to be su(2)$_1|_R$
$\times$ Z$_2|_L$. This result can also 
be derived using a Toulouse point approach \cite{phlechir98}.
As a whole, taking into
account of the contribution 
of the Majorana fermions $\xi^{0}_{R,L}$, 
the model (\ref{conintmod2}) displays critical 
properties characterized by a fixed point
with a su(2)$_{1}$ $\times$ Z$_2$ $\times$ Z$_2$
symmetry and a $c=2$ central charge. 
This emerging quantum criticality
can be interpreted 
as the criticality resulting from an effective $S=1/2$ AF Heisenberg
chain and two decoupled critical Ising models.
Frustration, introduced by the diagonal 
exchange interaction $J_{\times}$, shows up
in a non-trivial way by stabilizing 
a quantum critical point with central charge $c=2$.
This IR behavior differs from
the standard $c=1$ quantum criticality in unfrustrated 
three-leg spin ladder by the two gapless non-magnetic, singlet, 
degrees of freedom described by the Ising models.
The physical properties of the model at this 
$c=2$ critical point can be determined by 
a Toulouse point approach of 
the model (\ref{conintmod2}) \cite{phlechir98}.
The slowest spin-spin correlation functions 
of the model at the IR critical point correspond
to the staggered correlations between the spins of
the surface chains which decay with a universal
exponent $3/2$.
As a consequence, the low-temperature dependence
of the NMR relaxation rate scales 
as $1/T_1 \sim \sqrt{T}$ in contrast to
$1/T_1 \sim {\rm const}$ 
for  the spin-1/2 AF Heisenberg chain \cite{bookboso}.
The $c=2$ quantum critical point, induced by frustration,
describes thus a new universality class in spin ladders.
In addition, the Toulouse point analysis enables us
to investigate the effect of the neglected backscattering
term ${\vec n}_{2}
\cdot \left({\vec n}_1 +  {\vec n}_3 \right)$
of the model (\ref{conintmod1}) 
in the vicinity of the line ${\tilde g} =0$.
The main effect of this operator is to open 
a spectral gap in the Ising degrees of freedom 
of the $c=2$ fixed point but has no dramatic effect 
on the effective $S=1/2$ AF Heisenberg chain.
The $c=2$ IR fixed point is thus unstable
with respect to the interchain backscattering
term and the model (\ref{conintmod1}) will display 
IR critical properties governed by the $c=1$ fixed point
of the three-leg spin ladder or the $S=1/2$
AF Heisenberg spin chain.

Finally,
it is worth noting that the zero-temperature phase diagram of the three-leg
spin ladder with crossings (\ref{3legfrus})  has been recently
investigated by means of the DMRG approach \cite{wang02}.
Two critical phases have been found corresponding to the $S=3/2$ 
AF Heisenberg chain and the three-leg spin ladder.
Frustration, introduced through the diagonal interchain
interaction $J_{\times}$, induces a quantum phase
transition between these two critical phases
belonging to the standard $c=1$ universality class.
At large $J_{\times}$, the transition has been found to be
of first order. In the weak coupling regime,
the transition occurs 
at $J_{\perp} = 2 J_{\times}$ when $J_{\times} < 0.8 J_{\parallel}$
in full agreement with the field theoretical description (\ref{conintmod1})
of the model.
Extended DMRG calculations are required to fully 
characterize the nature of the phase boundary at $J_{\perp} = 2 J_{\times}$ 
in the weak coupling regime in particular to verify the 
chiral spin liquid behavior with $c=2$ criticality
proposed in Ref. \cite{phlechir98}.

\section{Concluding remarks}    

The suppression of magnetism by quantum fluctuations
in one dimension gives rise to a large variety of
zero-temperature spin liquid phases
with striking different physical properties. 
The canonical examples are the emerging quantum criticality
of the spin-1/2 AF Heisenberg chain, controlled 
by gapless elementary excitations
with fractional quantum numbers (spinons),
and the formation of an incompressible Haldane spin liquid
phase in the spin-1 case 
with optical $S=1$ magnon excitations. 
Several other types of incompressible spin liquid phases
can be stabilized with different dynamical or 
optical properties.
A distinct spin liquid behavior may be revealed in 
the form of the dynamical structure
factor with the presence or not of a sharp
spectral peak which directly probes the elementary
nature of the triplet excitation of the phase.
In this respect,
the staggered dimerized
phase of the two-leg spin ladder with a biquadratic
interchain exchange represents a non-Haldane spin liquid phase
even though a spectral gap is formed by quantum fluctuations. 
The main difference between this phase and the Haldane phase 
of the spin-1 Heisenberg chain stems from
the composite nature of the spin-flip $S=1$ excitation 
due to the presence of a two-particle threshold
in the dynamical structure factor instead of a sharp
magnon peak.
Two gapped spin liquid phases can also be distinguished
at the level of the topology of the short-range valence
bond description of their ground states.
A spin liquid phase may exhibit a ground-state degeneracy
depending on the nature of boundary conditions used.
In particular, for open boundary conditions, 
chain-end spin excitations can result from this 
ground-state degeneracy  leading
to well defined satellite peaks in the NMR profile of the system
doped with non-magnetic impurities.
An example of this topological distinction between
two gapped phases is provided by the two-leg spin ladder.
In the Haldane phase of the open 
two-leg ladder with a ferromagnetic rung interchain
$J_{\perp} < 0$, $S=1/2$ edge states are formed
whereas these chain-end excitations disappear in
the rung singlet phase of the ladder with $J_{\perp} > 0$.

Frustration, i.e. the impossibility to satisfy
simultaneously every pairwise interaction,
can induce additional types of spin liquid phases 
with exotic properties.
In the spin-1/2 case, one of the most striking effect
of frustration is the stabilization of a spontaneously
dimerized phase with massive deconfined spinon excitations.
Frustration represents thus a direct route to fractionalization.
The existence of these deconfined spinon excitations has
important consequences, as for instance, in the study of doping effects
of such a spin liquid phase with non-magnetic impurities.
On general grounds,
it is expected that
no free spin degrees of freedom are generated around non-magnetic 
impurities due to the presence of deconfined spinons \cite{sachdev01}.
In the case of the spontaneously dimerized phase of the 
spin-1/2 J$_1$-J$_2$  Heisenberg chain, 
the absence of induced free spin degrees of freedom or edge states
has been shown recently \cite{normand02}.
A second remarkable effect of frustration
is the onset of incommensurability in spin chains
whose classical ground state has a spiral structure.
In this respect, frustration is
a novel mechanism in one dimension
for generating incommensurate behavior as external
magnetic fields or Dzyaloshinskii-Moriya interaction.
An example of such a spin liquid phase with incommensurate
correlation is the quantum $J_1-J_2$ Heisenberg chain 
in the large next-nearest neighbor limit.
When the SU(2) symmetry of this model is explicitely 
broken, an incommensurate phase is produced by frustration 
with local non-zero spin currents, polarized along
the anisotropic axis, circulating around
triangular plaquettes.
This phase is the quantum signature of the classical
spiral phase induced by frustration.
Finally, a last effect of frustration, discussed in this review,
is the possible realization of a new 
type of emerging quantum criticality in spin chains and ladders.
In the continuum description,
the low-energy effective field theory of a frustrated 
spin chain is mainly governed by marginal interactions.
An extended IR critical behavior characterized by  
a CFT with central charge $c >1$ may result 
from the delicate balance between these marginal contributions.

Regarding perspectives, a natural question to raise is the existence
of a novel 1D spin liquid phase with unbroken SU(2) spin symmetry
which displays physical properties not described in this review.
In particular, an interesting possibility is 
the realization of a 1D version of the 
chiral spin liquid phase \cite{wen89} which 
breaks spontaneously the time-reversal symmetry.
In fact, such a phase has been identified recently
in the phase diagram of the two-leg spin ladder
with a four-spin cyclic exchange \cite{lauchli03,hikihara03}.
In the large  ring-exchange limit, 
a spin liquid phase characterized
by a non-zero scalar-chirality operator $\langle {\vec S}_{1,i} \cdot 
({\vec S}_{2,i} \wedge {\vec S}_{1,i+1})\rangle$,
breaking both parity and time-reversal symmetries,
has been found in DMRG calculations \cite{lauchli03,hikihara03}.
Using duality arguments, the authors of Ref. \cite{hikihara03}
have also exhibited several spin ladder models with 
exact ground-state which display spontaneous breaking
of the time-reversal symmetry.
Finally, a central issue is the possibility 
to stabilize a two-dimensional SU(2)-invariant spin liquid phase
with exotic properties starting from the one-dimensional limit.
Recently, 
a model of two-dimensional frustrated spin system
with SU(2) spin symmetry and strong spatially anisotropy
has been introduced by Nersesyan and Tsvelik \cite{nersesyan03}.
This model can also be interpreted as a collection of weakly 
coupled spin chains.
It has been shown that, in the limit of infinite number
of chains, this system displays a spin liquid phase
with massive deconfined fractional 
spin-1/2 excitations (spinons) \cite{nersesyan03}.
The existence of spinons in two-dimensional spin systems has also 
been found in a crossed-chain model \cite{starykh02}
which is a 2D anisotropic version of the 
pyrochlore lattice.
Remarkably enough, for a range of coupling constants \cite{starykh02,phsi02}, 
this system is a non-dimerized spin liquid with 
deconfined gapless spinons as elementary excitations.
This spin liquid phase, stabilized by frustration, 
is an example of a SU(2) version of a sliding 
Luttinger liquid phase \cite{slidinglutt00}.
We hope that other two-dimensional spin liquid phases with exotic
properties will be reported in the near future.

\section*{Acknowledgments}

We would like to thank D. Allen, P. Azaria, E. Boulat, B. Dou{\c c}ot, F. H. L. 
Essler, Th. Giamarchi, Th. Jolicoeur, A. A. Nersesyan, 
E. Orignac, T. T. Truong, and A. M. Tsvelik for hepful discussions
and their interest in this work.

\end{document}